\documentclass[11pt,a4paper]{article}
\pdfoutput=1
\usepackage{jheppub}
\usepackage{amsmath}
\usepackage{graphicx}
\usepackage{subcaption}
\usepackage{color,ulem}
\usepackage{makecell,multirow}

\allowdisplaybreaks

\title{Perturbativity constraints on $U(1)_{B-L}$ and left-right models and implications for heavy gauge boson searches}

\author[a]{Garv Chauhan,}
\author[a]{P. S. Bhupal Dev,}
\author[b]{Rabindra N. Mohapatra,}
\author[a,c]{Yongchao Zhang}
\affiliation[a]{Department of Physics and McDonnell Center for the Space Sciences,  Washington University, St. Louis, MO 63130, USA}
\affiliation[b]{Maryland Center for Fundamental Physics, Department of Physics, University of Maryland, College Park, MD 20742, USA}
\affiliation[c]{Center for High Energy Physics, Peking University, Beijing 100871, China}

\date{\today}

\begin{document}

\abstract{We derive perturbativity constraints on beyond standard model scenarios with extra gauge groups, such as $SU(2)$ or $U(1)$, whose generators contribute to the electric charge, and show that there are both upper and lower limits on the additional gauge couplings, from the requirement that the couplings remain perturbative up to the grand unification theory (GUT) scale. This leads to stringent constraints  on the masses of the corresponding gauge bosons and their collider phenomenology.
We specifically focus on the models based on $SU(2)_L\times U(1)_{I_{3R}} \times U(1)_{B-L}$ and the left-right symmetric models based on $SU(2)_L\times SU(2)_R\times U(1)_{B-L}$, and discuss  the implications of the perturbativity constraints for new gauge boson searches at current and future colliders. In particular, we find that the stringent flavor constraints in the scalar sector of left-right model set a lower bound on the right-handed scale $v_R \gtrsim 10$ TeV, if all the gauge and quartic couplings are to remain perturbative up to the GUT scale. This precludes the prospects of finding the $Z_R$ boson in the left-right model at the LHC, even in the high-luminosity phase, and leaves only a narrow window for the $W_R$ boson. A much broader allowed parameter space, with the right-handed scale $v_R$ up to $\simeq 87$ TeV, could be probed at the future 100 TeV collider.}

\maketitle

\section{Introduction}

The standard model (SM) of particle physics is astonishingly successful as a theory of particles and forces in the Universe. However, to account for several observations, such as the tiny neutrino masses, nature of the dark matter and  the origin of matter in the universe, the SM has to be extended to include new physics. Beyond the standard model (BSM) physics could of course be at any scale; however, from an experimental point of view, it is interesting if it is at the TeV scale so that it could be tested by current and planned experiments. Many TeV-scale BSM extensions proposed to remedy the above shortcomings of the SM introduce extended gauge groups,  such as extra $U(1)$ or $SU(2)\times U(1)$ groups at the TeV scale, which are usually derived from a higher symmetry group, such as
$SO(10)$~\cite{Georgi:1974my, Fritzsch:1974nn, Chanowitz:1977ye, Georgi:1979dq, Babu:1992ia} at the grand unification theory (GUT) scale. Such extensions broadly fall into two classes:
\begin{itemize}
\item [(i)] The generators of the extra gauge groups contribute to the electric charge~\cite{Marshak:1979fm, Mohapatra:1980qe}. Two widely discussed examples are (a) the models based on the gauge group $SU(2)_L\times U(1)_{I_{3R}}\times U(1)_{B-L}$~\cite{Deshpande:1979df, Galison:1983pa} and (b) the left-right symmetric model (LRSM) based on the gauge group $SU(2)_L\times SU(2)_{R}\times U(1)_{B-L}$~\cite{Pati:1974yy, Mohapatra:1974gc, Senjanovic:1975rk}, both of which  are useful and motivated in order to understand neutrino masses via the seesaw mechanism~\cite{seesaw1, seesaw2, seesaw3, seesaw4, seesaw5}.

\item [(ii)] The extra gauge groups do not contribute to electric charge. Examples of this class are the dark photon~\cite{Okun:1982xi, Holdom:1985ag, Foot:1991kb}, $U(1)_{B-L}$~\cite{Buchmuller:1991ce,Basso:2011hn}, and more generic $U(1)_X$~\cite{Appelquist:2002mw, Carena:2004xs, Das:2016zue, Oda:2017kwl} models, which have been discussed extensively in connection with dark matter~\cite{Alexander:2016aln} and collider signatures~\cite{Langacker:2008yv}. 
\end{itemize}

In both these classes of models, demanding that gauge couplings remain perturbative i.e. $g_i < \sqrt{4\pi}$ up to the GUT or Planck scale imposes severe constraints on the allowed values of the extra gauge couplings, as well as on the masses of the additional gauge bosons. In case (i), where the additional group generators contribute to the electric charge, we find both upper and lower limits on the gauge couplings, whereas in case (ii), where the additional gauge couplings are not related to the electric charge, we only get upper limits and no lower limits. In this paper, we only focus on the case (i) models and derive the perturbativity bounds on the gauge couplings $g_R$ and $g_{BL}$, corresponding to the $SU(2)_R$ (or $U(1)_{I_{3R}}$) and $U(1)_{B-L}$ gauge groups, respectively.

Our results have far-reaching implications for collider searches for extra gauge bosons. In particular, they have to be taken into consideration, while interpreting the current direct search constraints on the $W_R$~\cite{Sirunyan:2018pom, Aaboud:2018spl} and $Z_R$~\cite{ATLAS:2016cyf,CMS:2016abv} bosons from the Large Hadron Collider (LHC) data, or the prospects~\cite{Ferrari:2000sp, Dev:2015kca, Mitra:2016kov, Rizzo:2014xma, Ruiz:2017nip} at the High-Luminosity LHC (HL-LHC) and a future 100 TeV collider~\cite{Arkani-Hamed:2015vfh, Golling:2016gvc}. In particular, if the measured gauge couplings fall outside the limits derived from perturbativity up to the GUT (or Planck) scale, that would imply that there is new physics at the TeV  or intermediate scale which allows this to happen. That would have interesting implications for new BSM physics.

There is another important implication of our results for the LRSM. Due to the stringent flavor-changing neutral current (FCNC) constraints in the high-precision electroweak data such as $K_0 - \overline{K}_0$, $B_d - \overline{B}_d$ and $B_s - \overline{B}_s$ mixings~\cite{Tanabashi:2018oca}, the parity partner of the SM doublet scalar is required to be very heavy, i.e. $\gtrsim 10$ TeV~\cite{Ecker:1983uh, Zhang:2007da, Maiezza:2010ic, Bertolini:2014sua}.\footnote{Due to sizable hadronic uncertainties, the FCNC constraints on the heavy bidoublet scalars might go up to $\sim 25$ TeV.} Then one of the quartic couplings ($\alpha_3$) in the scalar potential [see Eq.~(\ref{eqn:potential})] is of order one, if the right-handed (RH) scale $v_R$ lies in the few-TeV range. As a result, the perturbativity of the quartic couplings up to the GUT scale imposes a lower bound on the $v_R$ scale, i.e. $v_R \gtrsim 10$ TeV. The renormalization group (RG) running of $\alpha_3$ and other quartic couplings involves the gauge couplings $g_R$ and/or $g_{BL}$.
Hence, the perturbativity constraints in the scalar sector of LRSM do not only narrow down significantly the allowed ranges for the gauge couplings $g_R$ and $g_{BL}$, but also supersede the current $W_R$ and $Z_R$ mass limits from the LHC, and even rule out the possibility of finding them at the HL-LHC (see Fig.~\ref{fig:LRSM3}). Therefore, if a heavy $W_R$ and/or $Z_R$ boson was to be found at the later stages of LHC, then either it does not belong to the LRSM, or the minimal LRSM has to be further extended at the TeV-scale or a higher intermediate scale, such that all the gauge, scalar and Yukawa couplings are perturbative up to the GUT scale.

Though we focus on the minimal $U(1)_{B-L}$ and LRSM gauge groups in this paper, the basic arguments and main results could easily be generalized to other gauge groups at the TeV scale, such as the $SU(3)_L \times U(1)_X$~\cite{Singer:1980sw, Frampton:1992wt, Pisano:1991ee, Foot:1992rh}, $SU(3)_L\times SU(3)_R\times U(1)_X$~\cite{Dias:2010vt, Reig:2016tuk, Reig:2016vtf, Borah:2017inr, Hati:2017aez}, and alternative left-right models with universal seesaw mechanism for the SM quarks and charged leptons~\cite{Berezhiani:1983hm, Chang:1986bp, Rajpoot:1987fca, Davidson:1987mh, Babu:1988mw, Babu:1989rb, Mohapatra:2014qva, Dev:2015vjd, Deppisch:2017vne, Patra:2017gak} or with a stable right-handed neutrino (RHN) dark matter~\cite{Dev:2016xcp, Dev:2016qeb}. However, our results do not apply to situations where the extra $U(1)$ groups emerge out of non-Abelian groups at an intermediate scale, since they will completely alter the ultraviolet (UV) behavior of the TeV scale $U(1)$ gauge couplings. String theories provide many  examples where extra $U(1)$'s persist till the string scale without necessarily being embedded in intermediate scale  non-Abelian groups~\cite{Langacker:2008yv, Hewett:1988xc}. However, if the extra TeV-scale gauge group in question is valid up to the GUT scale, where it gets embedded into a non-Abelian GUT group, 
 $SO(10)$~\cite{Georgi:1974my, Fritzsch:1974nn, Chanowitz:1977ye, Georgi:1979dq, Babu:1992ia} or higher rank groups, our results will be applicable and give useful information on the particle spectrum at the TeV scale.

This paper is organized as follows: In Section~\ref{sec:theoretical}, we sketch the basic theoretical arguments behind the perturbativity constraints on the gauge couplings that contribute to the electric charge. The application to the $SU(2)_L\times U(1)_{I_{3R}}\times U(1)_{B-L}$ gauge group is detailed in Section~\ref{sec:B-L}, along with the implications for searches of the heavy $Z_R$ boson and the $v_R$ scale at the LHC and future 100 TeV colliders. The analogous study for the LRSM gauge group $SU(2)_L\times SU(2)_R\times U(1)_{B-L}$ is performed in Section~\ref{sec:LRSM},  where we also include the phenomenological implications on the $W_R$, $Z_R$ searches at colliders. We conclude in Section~\ref{sec:conclusion}. The state-of-the-art two-loop RG equations for the gauge, quartic and Yukawa couplings in the LRSM are collected in Appendix~\ref{sec:RGE:LRSM}.

\section{Theoretical constraints}
\label{sec:theoretical}

Our basic strategy is as follows: In the SM, when the electroweak gauge group breaks down to the electromagnetic group, i.e. ${\cal G}_{\rm SM} \equiv SU(2)_L \times U(1)_Y \to U(1)_{\rm EM}$, the electric charge is given by
\begin{eqnarray}
Q \ = \ I_{3L}+\frac{Y}{2} \, ,
\end{eqnarray}
and we have the relation among the gauge couplings at the electroweak scale:
\begin{eqnarray}
\label{eqn:relation1}
\frac{1}{e^2} \ = \
\frac{1}{g_L^2} + \frac{1}{g_{Y}^2} \, ,
\end{eqnarray}
where $g_L, \ g_Y, \ e$  are  the gauge couplings for the $SU(2)_L$, $U(1)_Y$ and $U(1)_{\rm EM}$ gauge groups, respectively. Current experiments completely determine these coupling values at the electroweak scale~\cite{Tanabashi:2018oca}:
\begin{align}
e \ = \ 0.313 \pm 0.000022 \, , \qquad g_L \ = \ 0.652 \pm 0.00026 \, , \qquad g_Y \ = \ 0.357 \pm 0.000060 \, .
\end{align}
When the SM is extended in the gauge sector, to the gauge group  $SU(2)_L \times U(1)_{X} \times U(1)_{Z}$, such that the extra $U(1)_{X,\,Z}$'s both contribute to the electric charge, then the modified electric charge formula becomes
\begin{eqnarray}
Q \ = \ I_{3L}+I_X+I_Z \, .
\end{eqnarray}
This is also true if we replace one of the $U(1)_{X,Z}$'s with an $SU(2)$.  The corresponding relation involving the new gauge couplings become~\cite{Georgi:1977wk}:\footnote{Note that we are not using here any GUT normalizations for the $U(1)$ couplings in Eq.~(\ref{eqn:relation2}). For normalized couplings, the relation has to be altered accordingly.}
\begin{eqnarray}
\label{eqn:relation2}
\frac{1}{g_Y^2} \ = \
\frac{1}{g_X^2} + \frac{1}{g_{Z}^2} \,,
\end{eqnarray}
where $g_X$ and $g_Z$ are the gauge couplings for the $U(1)_X$ and $U(1)_Z$ gauge groups,  respectively. This relation holds at the scale $v_{X}$, where $U(1)_X\times U(1)_Z$ breaks down to the SM $U(1)_Y$ and correlates the couplings $g_{X,\,Z}$ to $g_Y$. Since the value of $g_Y$ is experimentally determined at any scale $v_X$ (with the appropriate SM RG evolution),  we must have $g_{X,\,Z}$ bounded from below in order to satisfy Eq.~(\ref{eqn:relation2}). On the other hand, requiring that the gauge couplings $g_{X,\,Z}$ remain perturbative till the GUT or Planck scale implies that $g_{X,\,Z}$ must also be bounded from above at any given scale $v_X$. In other words, the couplings $g_{X,\,Z}$ can neither be arbitrarily large nor arbitrarily small at the TeV-scale, allowing only a limited range for their values. This in turn constrains the mass of the extra heavy gauge boson $Z'$, which is given by $M_{Z^\prime}^2 \sim (g_X^2 + g_Z^2) v_{X}^2$. Clearly this has implications for the production of $Z'$ at colliders.

As an example, when the SM gauge group is extended to $SU(2)_L \times U(1)_{I_{3R}} \times U(1)_{B-L}$ as in Section~\ref{sec:B-L}, or to $SU(2)_L \times SU(2)_{R} \times U(1)_{B-L}$ as in Section~\ref{sec:LRSM}, the gauge couplings $g_{X,\,Z}$ are respectively $g_R$ and $g_{BL}$, and $v_R$ is the scale at which the extended gauge groups break down to the SM electroweak gauge group ${\cal G}_{\rm SM}$. 
Eq.~(\ref{eqn:relation2}) then implies  a {\it lower} bound on the coupling $g_R$~\cite{Dev:2016dja}:
\begin{eqnarray}
\label{eqn:bound}
r_g \ \equiv \ \frac{g_R}{g_L} \ > \ \tan\theta_w
\left( 1 - \frac{4\pi}{g_{BL}^2} \frac{\alpha_{\rm EM}}{\cos^2\theta_w} \right)^{-1/2} \,,
\end{eqnarray}
where $\theta_w \equiv {g_Y}/{g_L}$ is the weak mixing angle, and $\alpha_{\rm EM} \equiv e^2/4\pi$ is the fine-structure constant. For a phenomenologically-preferred TeV-scale $v_R$, if $g_{BL}$ is in the perturbative regime, we can set an {\it absolute} theoretical lower bound on $r_g > \tan\theta_w \simeq 0.55$~\cite{Dev:2016dja, Brehmer:2015cia}. One should note that the lower bound on $g_R$ depends on the $v_R$ scale. This is before requiring the perturbativity to persist up to the GUT or Planck scale. When perturbativity constraints are imposed, the lower limit on $g_R$ becomes more stringent, as we show below (see Figs.~\ref{fig:UoneBL_ZR}, \ref{fig:UoneBL_vR}, \ref{fig:LRSM3} and \ref{fig:LRSM4}). 


\section{$U(1)_{B-L} $ model}
\label{sec:B-L}

The first case we focus on is the $ SU(2)_L \times U(1)_{I_{3R}} \times U(1)_{B-L}$ model~\cite{Deshpande:1979df, Galison:1983pa} 
which possesses two BSM $U(1)$ gauge groups, i.e. $U(1)_{I_{3R}} \times U(1)_{B-L}$, which break down to the SM $U(1)_Y$ at a scale $v_R$. Labeling the gauge couplings for the groups $ U(1)_{I_{3R}}$ and $U(1)_{B-L}$ as $g_R$ and $g_{BL}$ respectively, we can set lower bounds on both $g_R$ and $g_{BL}$ at the $v_R$ scale from the coupling relation (\ref{eqn:relation2}), as well as upper bounds from the requirement that they remain perturbative up to the GUT scale, as argued in Section~\ref{sec:theoretical}.
\begin{table}[!t]
  \centering
  \caption{Particle content of the $SU(2)_L \times U(1)_{I_{3R}} \times U(1)_{B-L} $ model. }
  \label{tab:UoneBL}
  \begin{tabular}{cc c  c c}
  \hline\hline
    & $SU(2)_L$ & $U(1)_{I_{3R}}$ & $U(1)_{B-L}$ \\ \hline
  $Q $ &  $\mathbf{2}$ & 0 & $\frac13$ \\
  $u_R$ &  $\mathbf{1}$ & $+\frac12$ & $\frac13$ \\
  $d_R$ &  $\mathbf{1}$ & $-\frac12$ & $\frac13$ \\ \hline
  $L$ & $\mathbf{2}$ & 0 & $-1$ \\
  $N$ & $\mathbf{1}$ & $+\frac12$ & $-1$ \\
  $e_R$ & $\mathbf{1}$ & $-\frac12$ & $-1$ \\ \hline
  $H$  & $\mathbf{2}$ & $-\frac12$ & 0 \\
  $\Delta_R$  & $\mathbf{1}$ & $-1$ & 2 \\ \hline
  \end{tabular}
\end{table}

The particle content of this model~\cite{Dev:2017dui, Dev:2017xry} is presented in Table~\ref{tab:UoneBL}. Freedom from anomalies requires three RHNs which help to generate the tiny neutrino masses via the type-I seesaw mechanism~\cite{seesaw2}. In the scalar sector, one singlet $\Delta_R$ is used to beak the $U(1)$ groups and generate the RHN masses, while the doublet $H$ breaks the electroweak group, as in the SM. The one-loop renormalization group equations (RGEs) for the gauge couplings of the two $U(1)$'s are generated by the following $\beta$-functions:
\begin{align}
\label{eqn:UoneBL:beta}
16\pi^2 \beta (g_{I_{3R}}) \ & = \
\frac{9}{2} \, g_{I_{3R}}^3 \,, \\
16\pi^2 \beta (g_{BL}) \ & = \
3 \, g_{BL}^3 \,.
\label{eqn:UoneBL:beta1}
\end{align}
Note that we have not used GUT renormalized $g_{BL}$, since we are not considering coupling unification, but rather the implications for the heavy $Z_R$ boson searches at colliders. This model could be viewed in some sense as an ``effective'' TeV-scale theory of LRSM  with the $SU(2)_R$-breaking scale and the mass of the heavy $W_R$ boson at the GUT scale~\cite{Dev:2017dui, Dev:2017xry}. 
The $U(1)_{B-L}$ model discussed in this section could also be the TeV-scale effective theory of some GUT that contains $U(1)_{B-L}$ as a subgroup. 

\begin{figure}[t!]
  \centering
  \includegraphics[width=0.6\linewidth]{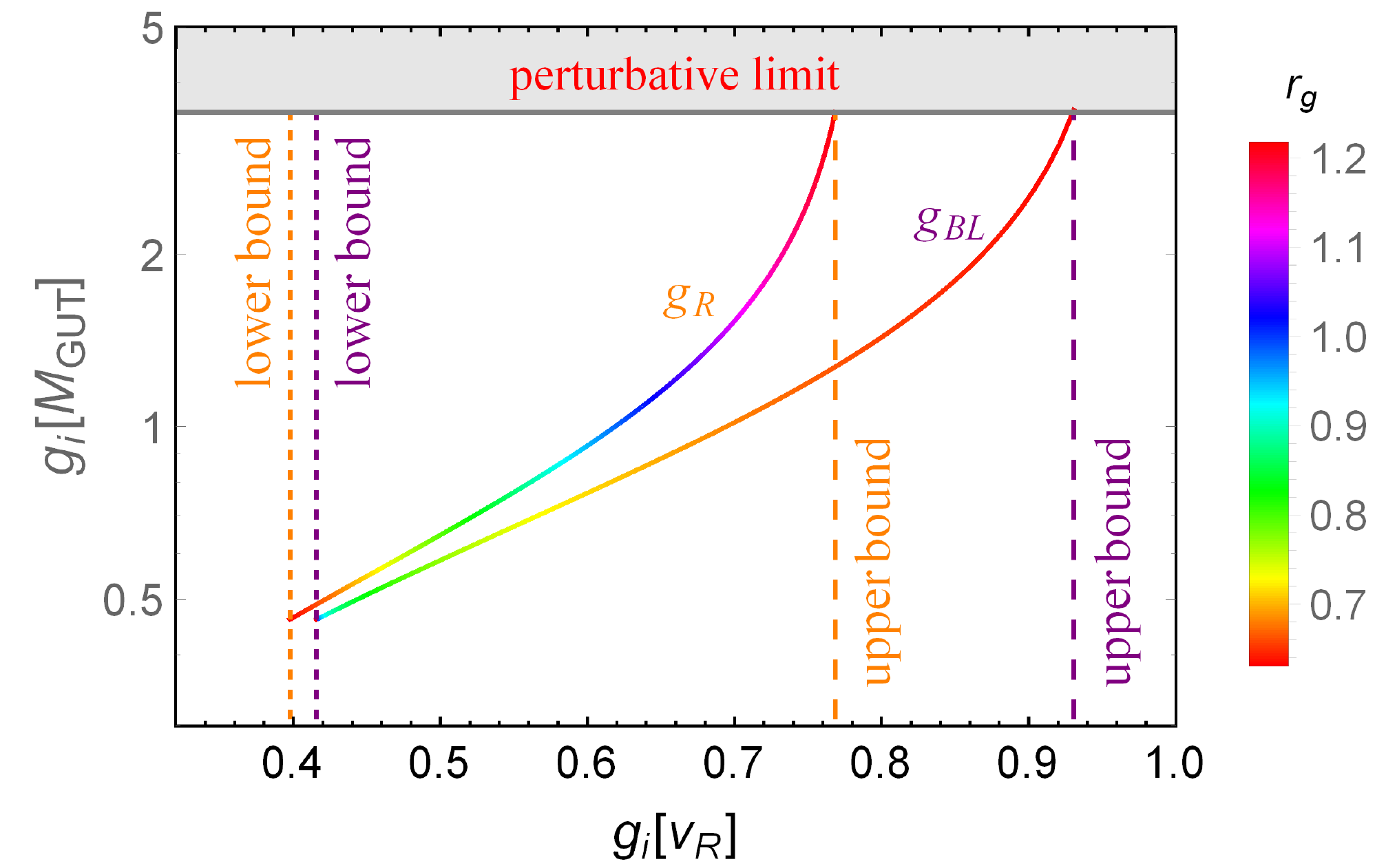}
  \caption{Correlation of $g_{R,\,BL} (v_R)$ and $g_{R,\,BL} (M_{\rm GUT})$ in the $U(1)_{B-L}$ model as functions of $r_g \equiv g_R / g_L$ at the $v_R$ scale (shown by the color coding). The horizontal shaded region is excluded by the perturbativity limit $g_{R,\,BL} < \sqrt{4\pi}$. The vertical dotted and dashed lines respectively denote the lower and upper limits on the gauge couplings. Here we have chosen $v_R=5$ TeV and $M_{\rm GUT}=10^{16}$ GeV.}
  \label{fig:UoneBL_g}
\end{figure}

\begin{figure}[!t]
  \centering
  \includegraphics[width=0.55\linewidth]{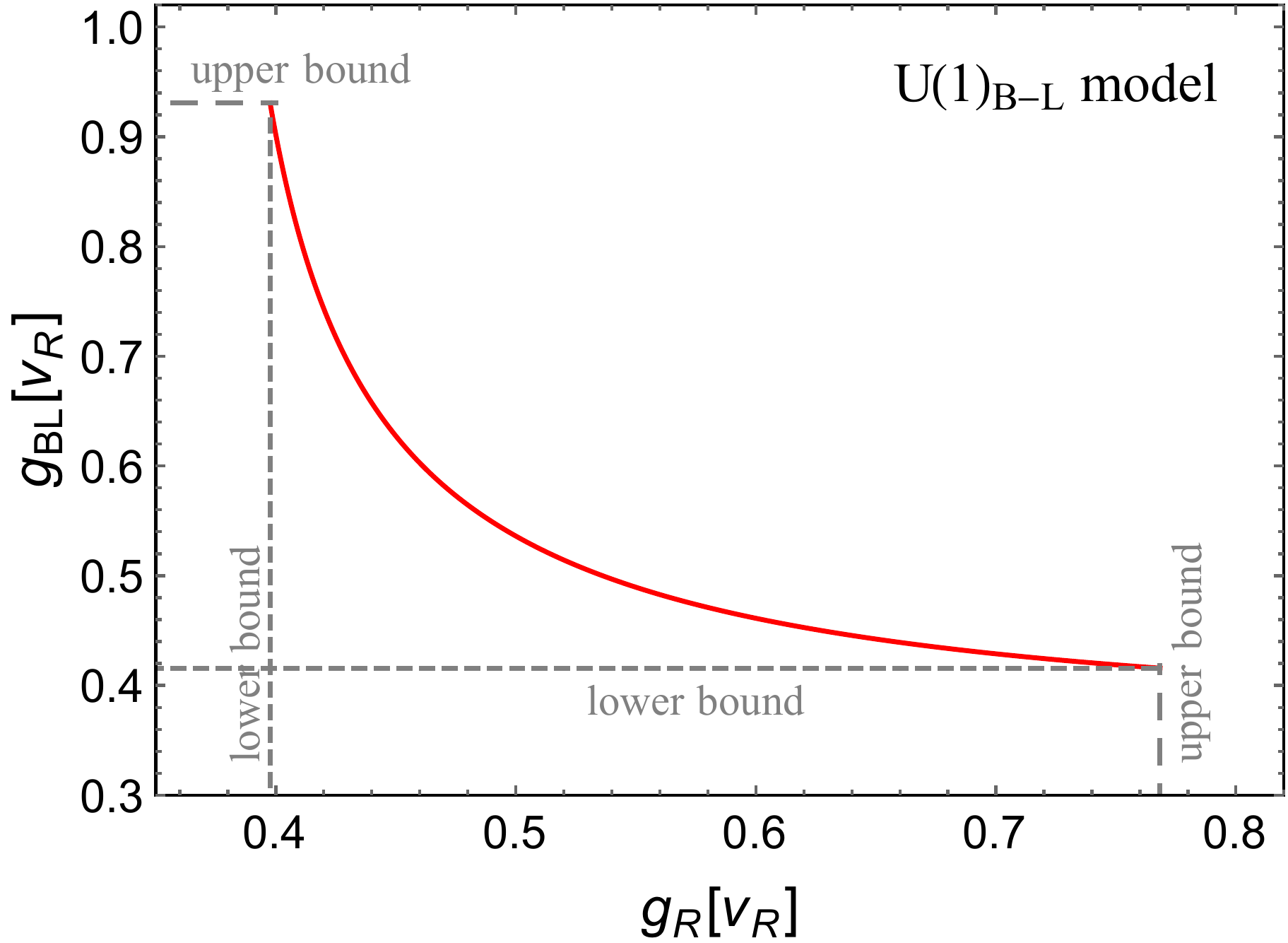}
  \caption{Correlation of $g_R$ and $g_{BL}$ at the scale $v_R = 5$ TeV (red line) and the lower and upper bounds on the couplings $g_R$ and $g_{BL}$, induced from the requirement of perturbativity up to the GUT scale in the $U(1)_{B-L}$ model.}
  \label{fig:UoneBL_g2}
\end{figure}

As an illustration, we set explicitly the RH scale $v_R = 5$ TeV, and run the SM coupling $g_Y$ from the electroweak scale $M_Z$ up to the $v_R$ scale, at which the couplings $g_R$ and $g_{BL}$ are related to $g_Y$ as in Eq.~(\ref{eqn:relation2}) and can be expressed as functions of the ratio $r_g \equiv g_R/g_L$. Then we evolve the two couplings $g_R$ and $g_Y$ from the $v_R$ scale up to the GUT scale, based on the $\beta$-functions in Eqs.~(\ref{eqn:UoneBL:beta}) and \eqref{eqn:UoneBL:beta1}. The correlations of $g_{R,\,BL}$ at the RH scale $v_R$ and GUT scale $M_{\rm GUT}=10^{16}$ GeV are presented in Fig.~\ref{fig:UoneBL_g}, as functions of the ratio $r_g$ at the $v_R$ scale (as shown by the color coding). The horizontal shaded region is excluded by the perturbativity limit $g_{R,\,BL} < \sqrt{4\pi}$. The vertical dashed lines denote the upper limits on the gauge couplings, requiring them to stay below the perturbativity limit up to the GUT scale. On the other hand, the vertical dotted lines denote the lower limits on the gauge couplings, obtained from Eq.~(\ref{eqn:relation2}), which implies there is only one degree of freedom in the $U(1)_{B-L}$ model, and the values of $g_R$ and $g_{BL}$ are correlated at the scale $v_R$, as shown by the red curve in Fig.~\ref{fig:UoneBL_g2}. In other words, a lower bound on $g_R$ corresponds to an upper bound on $g_{BL}$, and vice versa. Numerically, the gauge couplings are found to be constrained to lie within a narrow window
\begin{eqnarray}
\label{eqn:window2}
0.398  <  g_R < 0.768 \;\;   \text{and} \;\;
0.416  <  g_{BL}  < 0.931 \,,  \quad \text{with} \quad 0.631  <  r_g  < 1.218
\end{eqnarray}
at the $v_R$ scale, as shown in Fig.~\ref{fig:UoneBL_g}.

The perturbativity constraints on the gauge couplings $g_R$ and $g_{BL}$ at the $v_R$ scale have profound implications for the searches of the heavy $Z_R$ boson, whose mass is given by
\begin{eqnarray}
M_{Z_R}^2 \ \simeq \ 2(g_R^2 + g_{BL}^2)  \, v_R^2 \,.
\label{eq:mzr}
\end{eqnarray}
The $Z_R$ couplings to the chiral fermions $f_{L,R}$ are respectively~\cite{Dev:2017xry}
\begin{eqnarray}
\label{eqn:ZR1}
g_{Z_R f_L f_L} & \ = \ & \frac{e}{\cos\theta_w} \,
(I_{3,f} - Q_f) \,
\frac{\sin\phi}{\cos\phi} \,, \\
\label{eqn:ZR2}
g_{Z_R f_R f_R} & \ = \ & \frac{e}{\cos\theta_w} \,
(I_{3,f} - Q_f \sin^2\phi) \,
\frac{1}{\sin\phi \cos\phi} \,
\end{eqnarray}
with $Q_f$ the electric charge of fermion $f$,  $I_{3,f}$ the third-component of isospin of that particle, and $\tan\phi \equiv g_{BL} / g_R$ the RH gauge mixing angle.

\begin{table}[!t]
  \centering
  \caption{The lower bounds on the $Z_R$ boson mass $M_{Z_R}$ and the $v_R$ scale in the $U(1)_{B-L}$ model from the current LHC13 data~\cite{ATLAS:2016cyf, CMS:2016abv} and the prospects at the HL-LHC 14 TeV with an integrated luminosity of 3000 fb$^{-1}$~\cite{Diener:2010sy,Godfrey:2013eta} and future 100 TeV collider FCC-hh with a luminosity of 30 ab$^{-1}$~\cite{Godfrey:2013eta, Rizzo:2014xma}. The range in each case corresponds to the allowed range of $r_g$ from perturbativity constraints, as given in Eq.~\eqref{eqn:window2}.}
  \label{tab:UoneBL_limit}
  \begin{tabular}{l c c}
  \hline\hline
  collider & $M_{Z_R}$ [TeV] & $v_R$ [TeV]  \\ \hline
  LHC13 & $[3.6,\, 4.2]$ & $[3.02,\, 3.57]$  \\
  HL-LHC & $[6.0,\, 6.6]$ & $[4.60,\, 5.82]$ \\
  FCC-hh & $[27.9, 31.8]$ & $[19.9,\, 26.8]$ \\ \hline
  \end{tabular}
\end{table}

\begin{figure}[t]
  \centering
  \includegraphics[width=0.55\linewidth]{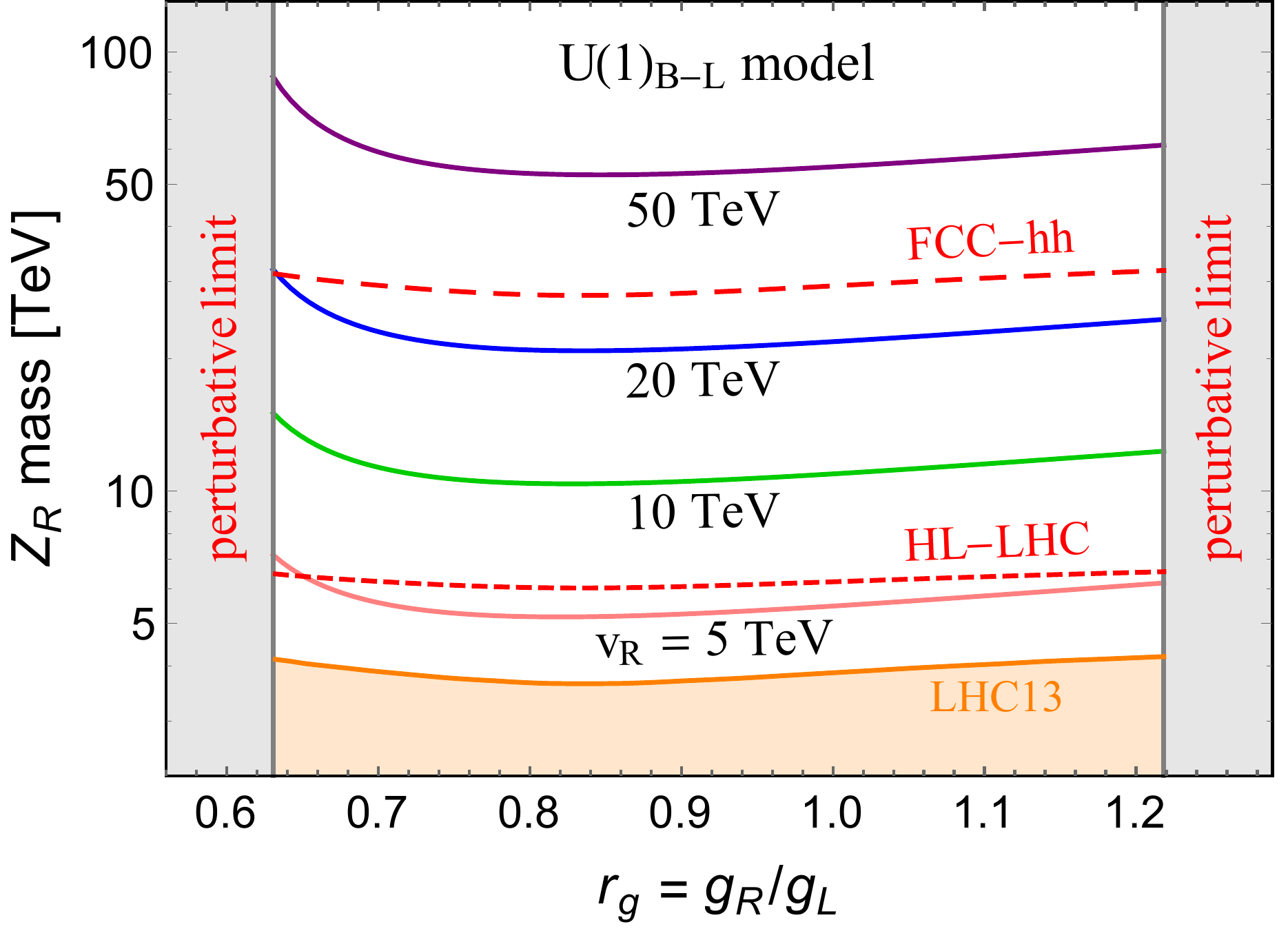}
  \caption{Current LHC13 constraints on $Z_R$ mass in the $U(1)_{B-L}$ model (shaded orange) as function of $r_g = g_R/g_L$, and future prospects at the HL-LHC 14 TeV with an integrated luminosity of 3000 fb$^{-1}$ (short-dashed red) and the 100 TeV collider FCC-hh with a luminosity of 30 ab$^{-1}$ (long-dashed red). The vertical shaded regions are excluded by the perturbativity constraints given in Eq.~\eqref{eqn:window2}. The pink, green, blue and purple contours show the variation of the $Z_R$ mass with respect to $r_g$, with the RH scale $v_R = 5$, 10, 20, 50 TeV, respectively. }
  \label{fig:UoneBL_ZR}
\end{figure}

For a TeV-scale $v_R$, the $Z_R$ mass is stringently constrained by the dilepton data $pp \to Z_R \to \ell^+ \ell^-$ (with $\ell = e,\, \mu$) at the LHC~\cite{Patra:2015bga, Lindner:2016lpp}. For a sequential $Z'$ boson, the current mass limit is 4.05 TeV at the 95\% confidence level (CL)~\cite{ATLAS:2016cyf, CMS:2016abv}. The dilepton prospects of a sequential $Z'$ boson have also been estimated at the HL-LHC~\cite{Diener:2010sy,Godfrey:2013eta} and future 100 TeV colliders~\cite{Godfrey:2013eta, Rizzo:2014xma}, which are respectively 6.4 TeV and 30.7 TeV, for an integrated luminosity of 3000 fb$^{-1}$. Given a luminosity of 10 times larger at  the 100 TeV collider, the dilepton prospects could be significantly enhanced, up to 43.7 TeV. The production cross section $\sigma (pp \to Z_R \to \ell^+ \ell^-)$ in the $U(1)_{B-L}$ model can be obtained by rescaling that of a sequential heavy $Z'$ boson, as function of $r_g = g_R/g_L$. The rescaled current mass limit and the expected limits at the HL-LHC and the future 100 TeV collider FCC-hh are presented in Fig.~\ref{fig:UoneBL_ZR}, as a function of $r_g$. The $Z_R$ mass contours for $v_R = 5$, 10, 20 and 50 TeV are also shown in Fig.~\ref{fig:UoneBL_ZR} in the colors of pink, green, blue and purple, respectively. The vertical shaded regions are excluded by the perturbativity constraints given in Eq.~\eqref{eqn:window2}.\footnote{When the $v_R$ scale changes from 5 TeV, the perturbative constraints on $r_g$ in Fig.~\ref{fig:UoneBL_ZR} will change accordingly from those given below Eq.~\eqref{eqn:window2}. However, this change is negligible for $v_R$ up to 50 TeV.} Fig.~\ref{fig:UoneBL_ZR} implies that the LHC13 lower limits, as well as the future HL-LHC and FCC-hh limits, on $Z_R$ boson mass are in a narrow range, depending on the allowed values of $r_g$, as shown in Table~\ref{tab:UoneBL_limit}. Thus, the perturbativity constraints restrict the accessible range of $M_{Z_R}$ up to 6.6 TeV at the HL-LHC and 31.8 TeV at the FCC-hh.
For the purpose of concreteness, we have assumed the decay $Z_R \to N_i N_i$ is open, such that the ${\rm BR} (Z_R \to \ell^+ \ell^-)$ is slightly smaller than the case without the decaying of $Z_R$ into RHNs and the dilepton limits in Fig.~\ref{fig:UoneBL_ZR} are comparatively more conservative~\cite{Dev:2016xcp}.

\begin{figure}[ht!]
  \centering
  \includegraphics[width=0.55\linewidth]{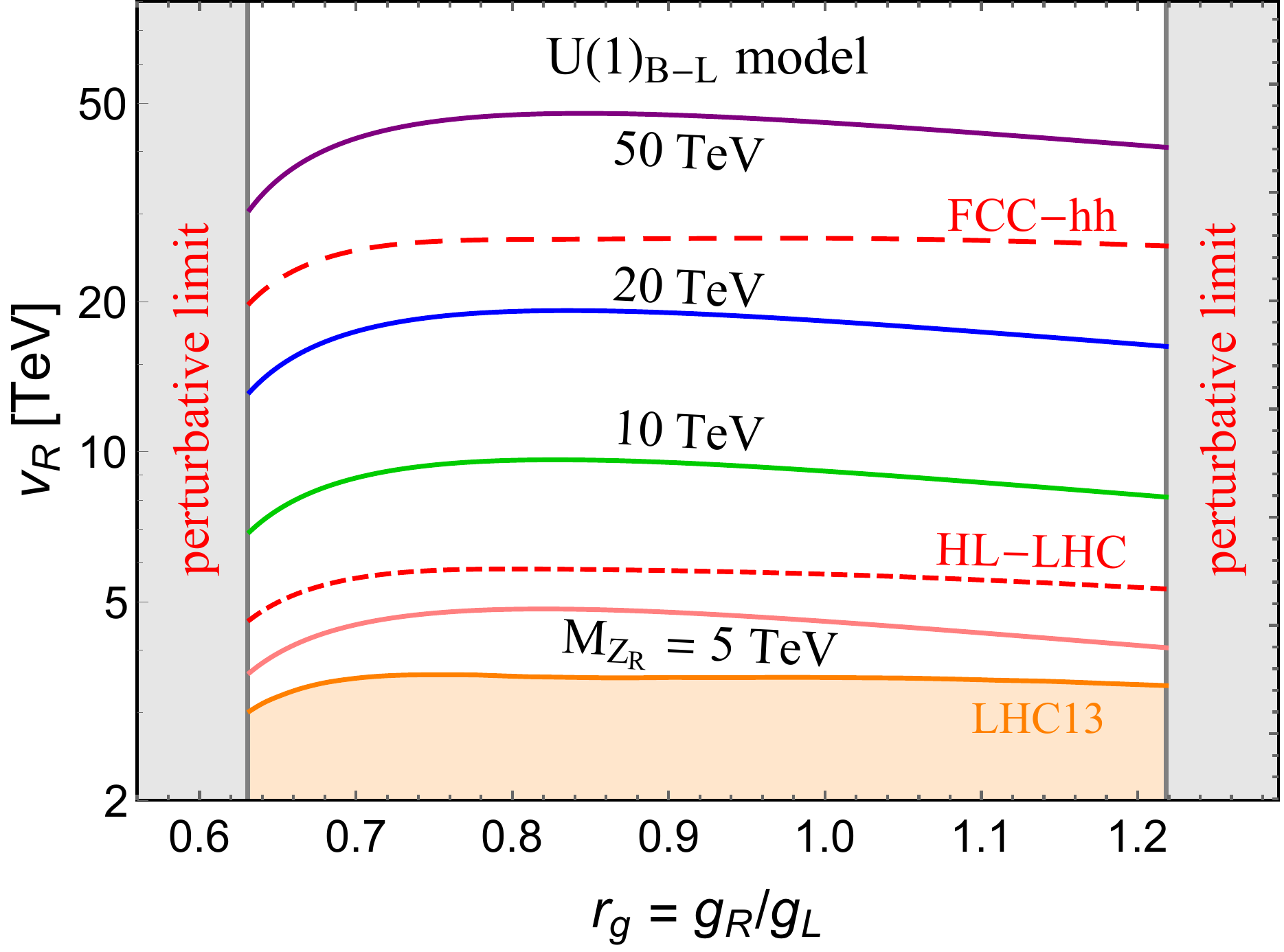}
  \caption{Current lower bound on the $v_R$ scale in the $U(1)_{B-L}$ model,  as functions of $r_g = g_R/g_L$, from the searches of $Z_R$ in the dilepton channel at LHC 13 TeV (shaded orange), as well as the future limit from HL-LHC 14 TeV with an integrated luminosity of 3000 fb$^{-1}$ (short-dashed red) and the 100 TeV collider FCC-hh with a luminosity of 30 ab$^{-1}$ (long-dashed red). The vertical shaded regions are excluded by the perturbativity constraints given in Eq.~\eqref{eqn:window2}. The pink, green, blue and purple contours show the variation of $v_R$ with respect to $r_g$, with the $Z_R$ mass $M_{Z_R} = 5$, 10, 20, 50 TeV, respectively. }
  \label{fig:UoneBL_vR}
\end{figure}

The dilepton constraints on the $Z_R$ mass can be traded for the constraints on $v_R$ scale using Eq.~\eqref{eq:mzr}. This is shown in Fig.~\ref{fig:UoneBL_vR} and Table~\ref{tab:UoneBL_limit}. The pink, green, blue and purple contours here show the variation of $v_R$ with respect to $r_g$, with fixed $Z_R$ mass of $M_{Z_R} = 5$, 10, 20, 50 TeV, respectively. The perturbativity constraints given in Eq.~\eqref{eqn:window2} restrict the accessible range of $v_R$ up to 5.8 TeV at the HL-LHC and 26.8 TeV at the FCC-hh.

\section{The minimal left-right symmetric model}
\label{sec:LRSM}

We now consider the TeV-scale LRSM based on the gauge group $ SU(2)_L \times SU(2)_{R} \times U(1)_{B-L} $~\cite{Pati:1974yy, Mohapatra:1974gc, Senjanovic:1975rk}. Original aim of this model was to explain the asymmetric chiral structure of electroweak interactions in the SM. It was subsequently pointed out that it could account for the observed small neutrino masses via the type-I~\cite{seesaw1, seesaw2, seesaw3, seesaw4, seesaw5} and/or type-II~\cite{Magg:1980ut, Schechter:1980gr, Mohapatra:1980yp, Lazarides:1980nt, Konetschny:1977bn, Cheng:1980qt} seesaw mechanisms. In the ``canonical'' version of LRSM, it is always assumed that the gauge coupling $g_R = g_L$ and the scalar content of the LRSM consists of one bidoublet $\Phi$ and the left-handed ($\Delta_L$) and right-handed ($\Delta_R$) triplets. As long as the RH scale $v_R$ is at the few-TeV range, the values of $g_R$ and $g_{BL}$ and their RG running up to the GUT scale are almost fixed, at least at the one-loop level. However, the coupling $g_R$ might be different from $g_L$, which generates very rich phenomenology in the LRSM, see e.g.~\cite{Dev:2016dja, Dev:2016xcp, Deppisch:2014zta, Brehmer:2015cia, Dev:2015pga}. Moreover, a free $g_R$ not necessarily equal to  $g_L$ makes it possible to investigate the whole parameter space of perturbative constraints in the LRSM. In addition, the parity and $SU(2)_R$ breaking scales might also be different such that the  left-handed triplet $\Delta_L$ decouples from the TeV-scale physics~\cite{Chang:1983fu}. This also helps to avoid the unacceptably large type-II seesaw contribution to the neutrino masses and/or fine-tuning in the scalar sector. Based on these arguments, we will not consider the $\Delta_L$ field in the low-energy LRSM. The matter content and the scalar fields in the minimal LRSM are collected in Table~\ref{tab:LRSM}. Three RHNs $N_{1,2,3}$ have been naturally introduced to form the RH lepton doublets $\psi_R$ and accommodate the type-I seesaw mechanism. The perturbative constraints from the gauge and scalar sectors follow in the next two subsections.

\begin{table}[!t]
  \centering
  \caption{Particle content of the minimal LRSM based on the gauge group $SU(2)_L \times SU(2)_{R} \times U(1)_{B-L} $. }
  \label{tab:LRSM}
  \begin{tabular}{l c c c}
  \hline\hline
  &  $SU(2)_L$ & $SU(2)_{R}$ & $U(1)_{B-L}$\\ \hline
  $Q_L \equiv \begin{pmatrix} u_L \\ d_L \end{pmatrix}$ & $\mathbf{2}$ & $\mathbf{1}$ & $\frac13$ \\
  $Q_R \equiv \begin{pmatrix} u_R \\ d_R \end{pmatrix}$  & $\mathbf{1}$ & $\mathbf{2}$ & $\frac13$ \\ \hline
  $\psi_L \equiv \begin{pmatrix} \nu_L \\ e_L \end{pmatrix}$ & $\mathbf{2}$ &$\mathbf{1}$ & $-1$ \\
  $\psi_R \equiv \begin{pmatrix} N \\ e_R \end{pmatrix}$ & $\mathbf{1}$ &$\mathbf{2}$ & $-1$ \\  \hline
  $\Phi = \left(\begin{matrix}\phi^0_1 & \phi^+_2\\\phi^-_1 & \phi^0_2\end{matrix}\right)$ & $\mathbf{2}$ & $\mathbf{2}$ & 0 \\
  $\Delta_R = \left(\begin{matrix} \frac{1}{\sqrt2} \Delta^+_R & \Delta^{++}_R \\ \Delta^0_R & - \frac{1}{\sqrt2} \Delta^+_R \end{matrix}\right)$  & $\mathbf{1}$ & $\mathbf{3}$ & 2 \\ \hline
  \end{tabular}
\end{table}

\subsection{Perturbativity constraints from the gauge sector}
\label{sec:gauge}

The perturbativity limits in the gauge sector are conceptually similar to the $U(1)_{B-L}$ model in Section~\ref{sec:B-L}; the difference is mainly due to the $\beta$-function coefficients, which in this case are given by
\begin{align}
\label{eqn:LRSMBL:beta}
16\pi^2 \beta (g_R) \ & = \
-\frac{7}{3} \, g_R^3 \,, \\
16\pi^2 \beta (g_{BL}) \ & = \
\frac{11}{3} \, g_{BL}^3 \,.
\label{eqn:LRSMBL:beta1}
\end{align}
Note the change in sign for $\beta(g_R)$, as compared to Eq.~\eqref{eqn:UoneBL:beta}, which is due to the non-Abelian nature of $SU(2)_R$. For completeness, we have also computed the two-loop RGEs using the code {\tt PyR@TE}~\cite{Lyonnet:2013dna, Lyonnet:2015jca} and list them in Appendix~\ref{sec:RGE:LRSM}, although it turns out that the two-loop corrections change the results only by a few per cent, as compared to the one-loop results presented here.

\begin{figure}[!t]
  \centering
  \includegraphics[width=0.58\linewidth]{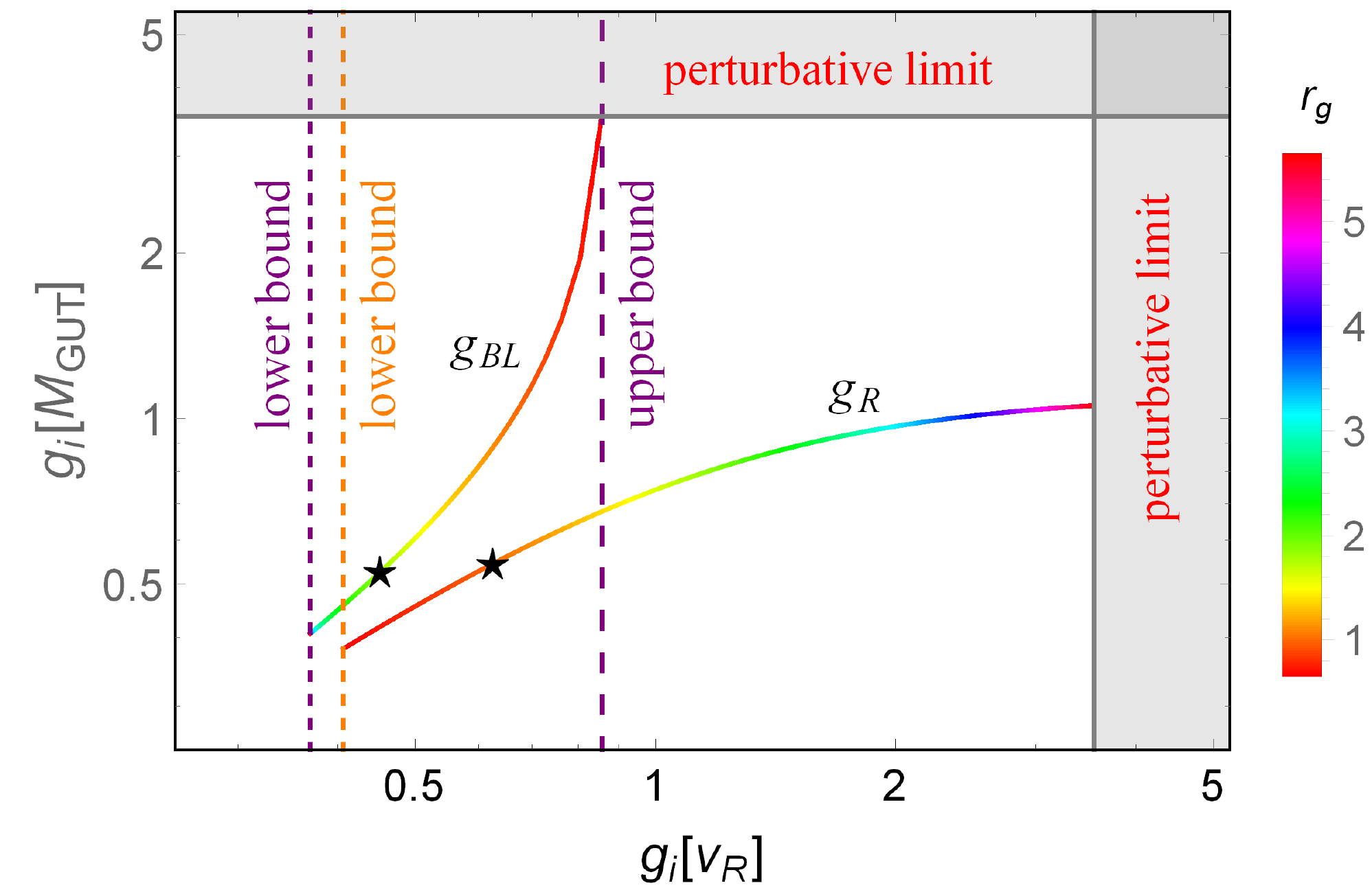}
  \caption{Correlation of $g_{R,\,BL} (v_R)$ and $g_{R,\,BL} (M_{\rm GUT})$ in the minimal LRSM as functions of $r_g = g_R/g_L$ at the $v_R$ scale (as shown by the color coding). The shaded regions are excluded by the perturbativity limits $g_{R,\,BL} < \sqrt{4\pi}$. The two stars correspond to the special case $g_R = g_L$ at the $v_R$ scale. The vertical dotted and dashed lines respectively denote the lower and upper limits on the gauge couplings. Here we have chosen $v_R=10$ TeV and $M_{\rm GUT}=10^{16}$ GeV.}
  \label{fig:LRSM1}
\end{figure}

\begin{figure}[!t]
  \centering
  \includegraphics[width=0.55\linewidth]{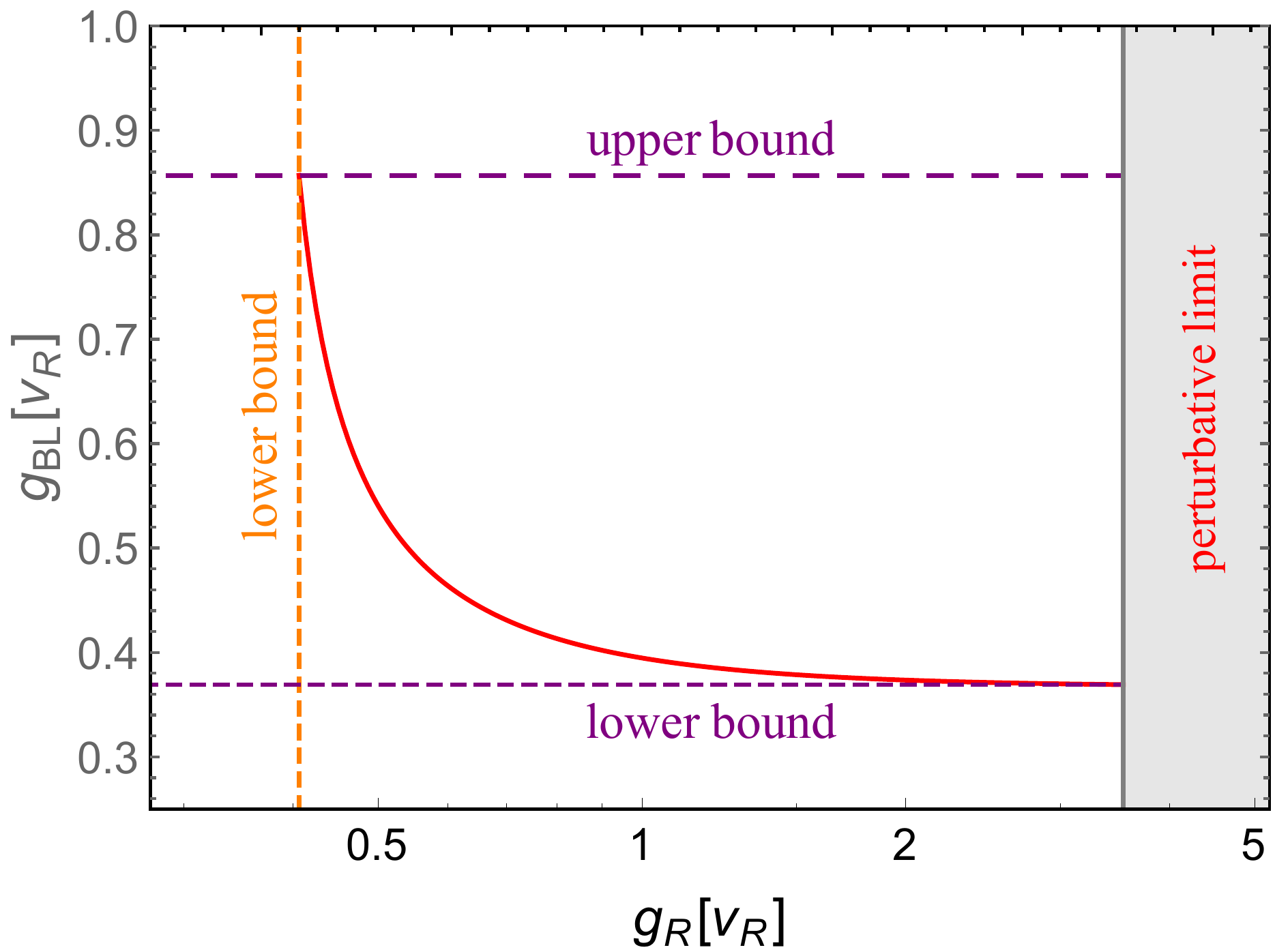}
  \caption{Correlation of $g_R$ and $g_{BL}$ in the minimal LRSM at the scale $v_R = 10$ TeV (red curve), along with the lower and upper bounds on the couplings $g_{R,\,BL}$, induced from the requirement of perturbativity up to the GUT scale. The shaded region is excluded by the perturbativity limit $g_{R} < \sqrt{4\pi}$.}
  \label{fig:LRSM2}
\end{figure}

\begin{figure}[!t]
  \centering
  \includegraphics[width=0.495\linewidth]{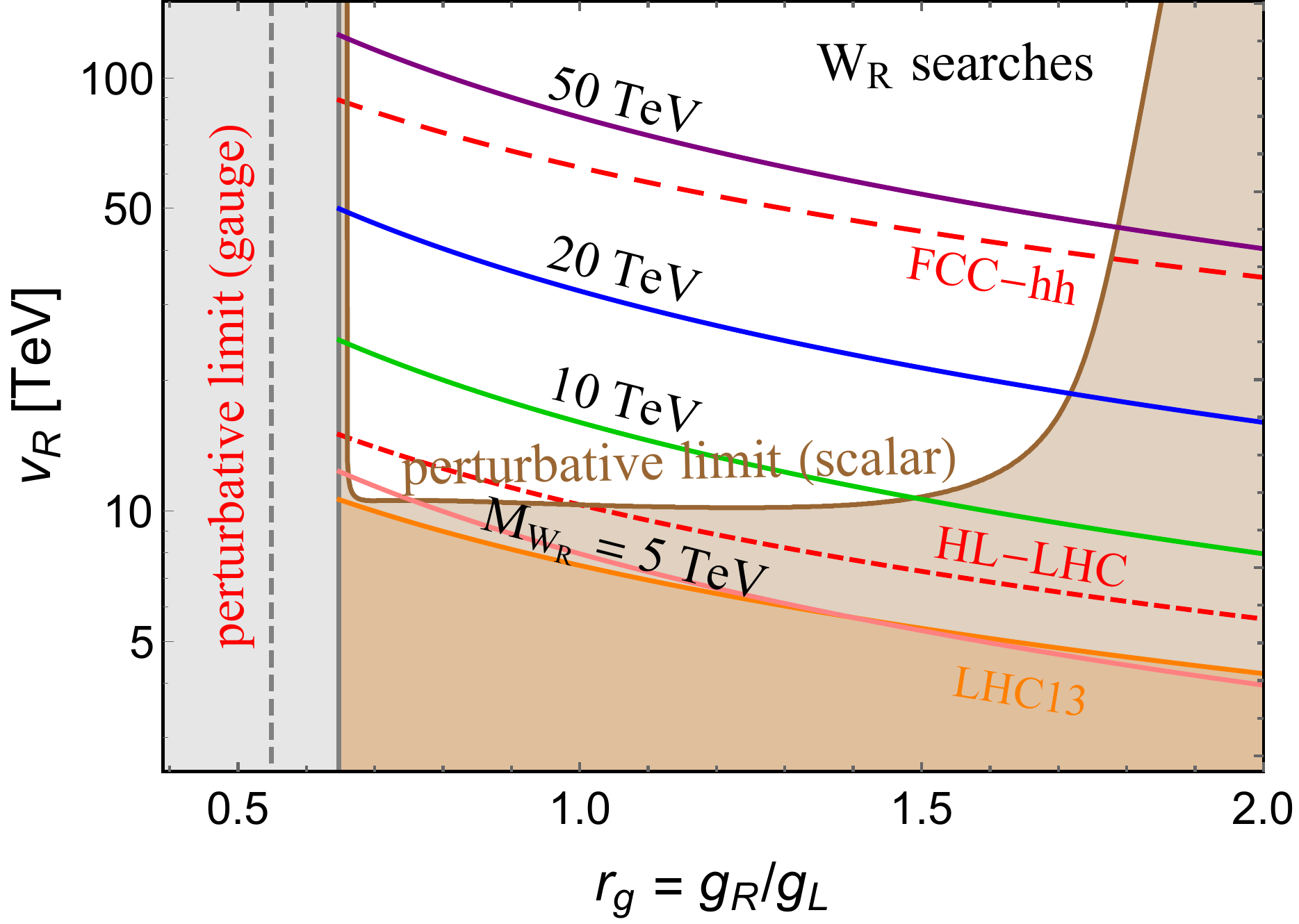}
  \includegraphics[width=0.49\linewidth]{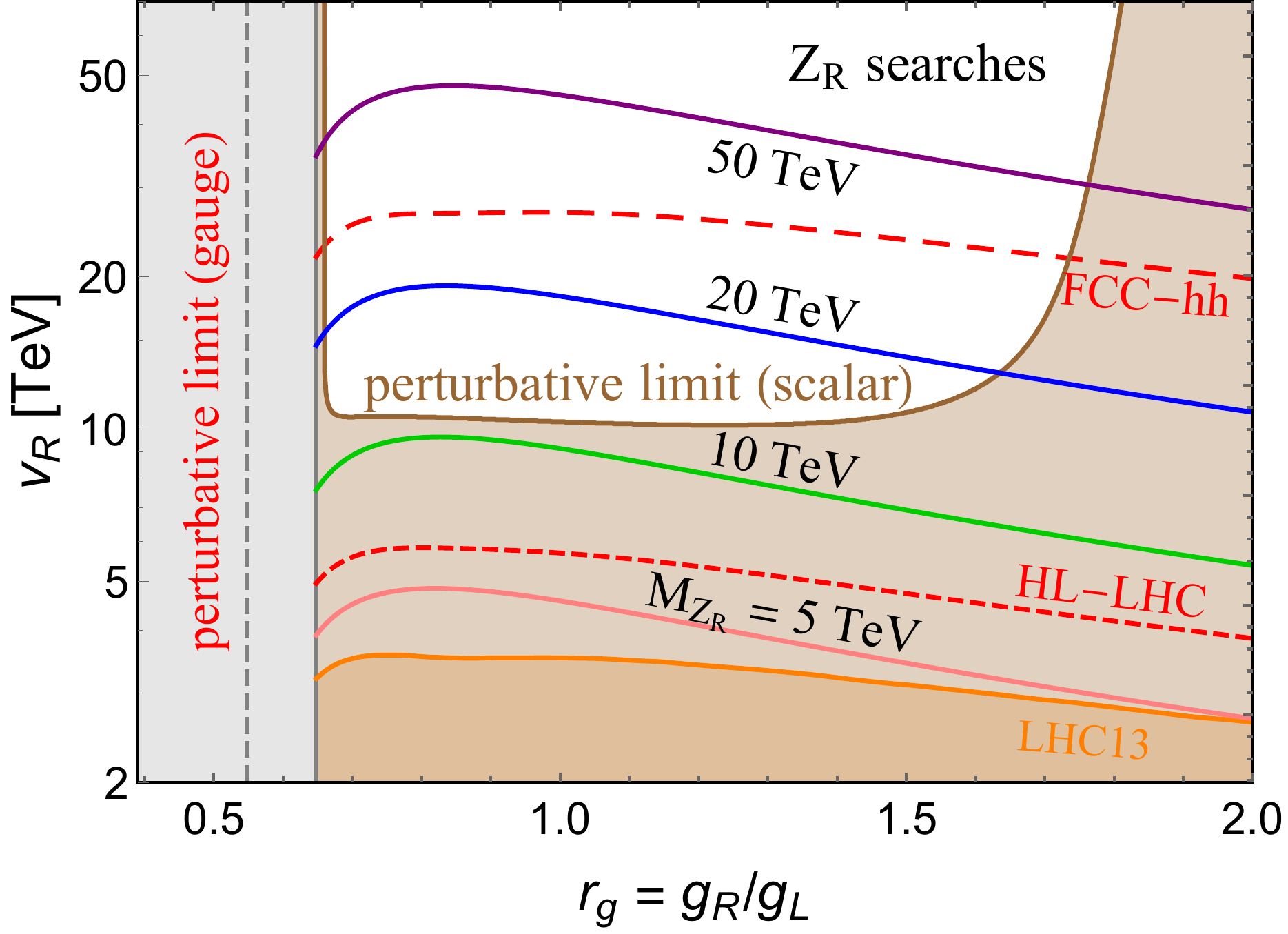}
  \caption{Lower bounds on the $v_R$ scale in the minimal LRSM,  as functions of $r_g$, from the direct searches of $W_R$ and $Z_R$ bosons at LHC 13 TeV (shaded orange), and future prospects at the HL-LHC 14 TeV with an integrated luminosity of 3000 fb$^{-1}$ (short-dashed red) and the 100 TeV collider FCC-hh with a luminosity of 30 ab$^{-1}$ (long-dashed red). The shaded gray regions are excluded by the perturbativity constraints up to the GUT scale, with the vertical dashed line corresponding to the absolute theoretical bound in Eq.~(\ref{eqn:bound}). The shaded brown regions are excluded by the perturbativity limits from the scalar sector, discussed in Section~\ref{sec:scalar}.}
  \label{fig:LRSM4}
\end{figure}

As the RH scale $v_R = 5$ TeV (chosen in Section~\ref{sec:B-L}) is in tension with the stringent constraints from the scalar sector in LRSM (see Section~\ref{sec:scalar} and Fig.~\ref{fig:LRSM4}), we set $v_R = 10$ TeV as an illustrative example to evaluate the perturbative constraints on the gauge couplings $g_R$ and $g_{BL}$. In fact, as long as the $v_R$ scale is at the ballpark of few-TeV, the changes in the running of $g_R$ and $g_{BL}$ are mainly due to the initial values of  $g_{R,\,BL}$ at the $v_R$ scale, and are negligibly small. As in the $U(1)_{B-L}$ model in Section~\ref{sec:B-L}, the couplings $g_R$ and $g_{BL}$ are both functions of the ratio $r_g = g_R/g_L$. The correlations of $g_{R,\,BL}$ at the $v_R$ scale and the GUT scale are presented in Fig.~\ref{fig:LRSM1}, as functions of $r_g$ (as shown by the color coding). The two stars in Fig.~\ref{fig:LRSM1} correspond to the special case $g_R = g_L$ at the $v_R$ scale. As a result of non-Abelian nature of the $SU(2)_R$ group, $g_R$ is asymptotically free, i.e. it becomes smaller at higher energy scales. Thus $g_R$ could go up to the perturbativity limit of $\sqrt{4\pi}$ at the $v_R$ scale (without considering the perturbativity limits from the scalar sector for the moment), which is very different from the $U(1)_{B-L}$ model, where the $g_R$ value is much more restricted at the $v_R$ scale [cf.~Fig.~\ref{fig:UoneBL_g}].  The allowed ranges of the gauge couplings in the minimal LRSM are
\begin{eqnarray}
\label{eqn:window3}
0.406 < g_R < \sqrt{4\pi} \;\; \text{and} \;\;
0.369 < g_{BL} < 0.857 \,, \quad \text{with} \quad
0.648 < r_g < 5.65
\end{eqnarray}
at the scale $v_R$, which is clearly shown in the correlation plot of $g_R$ and $g_{BL}$ in Fig.~\ref{fig:LRSM2}.

\begin{figure}[!t]
  \centering
  \includegraphics[width=0.49\linewidth]{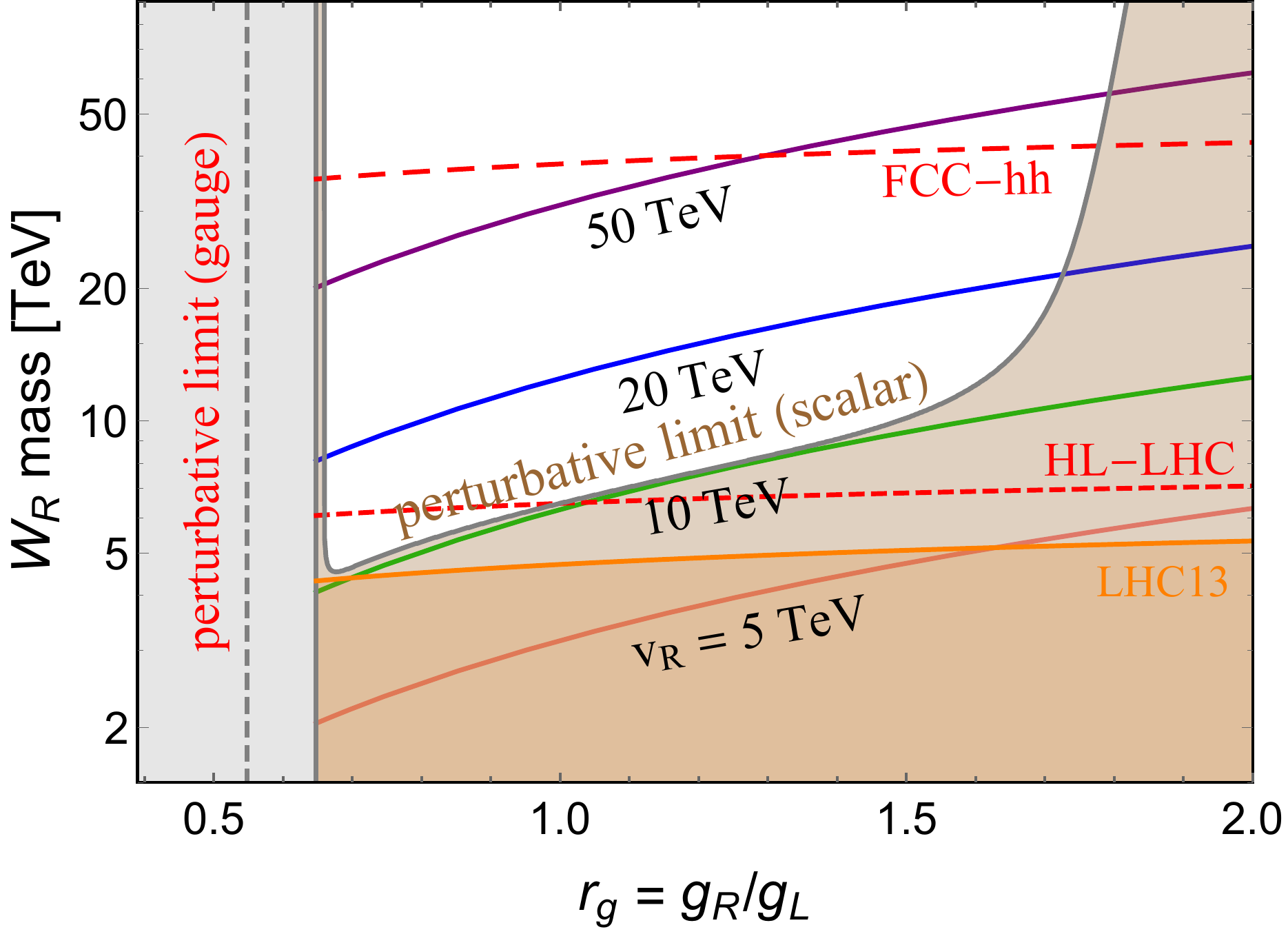}
  \includegraphics[width=0.49\linewidth]{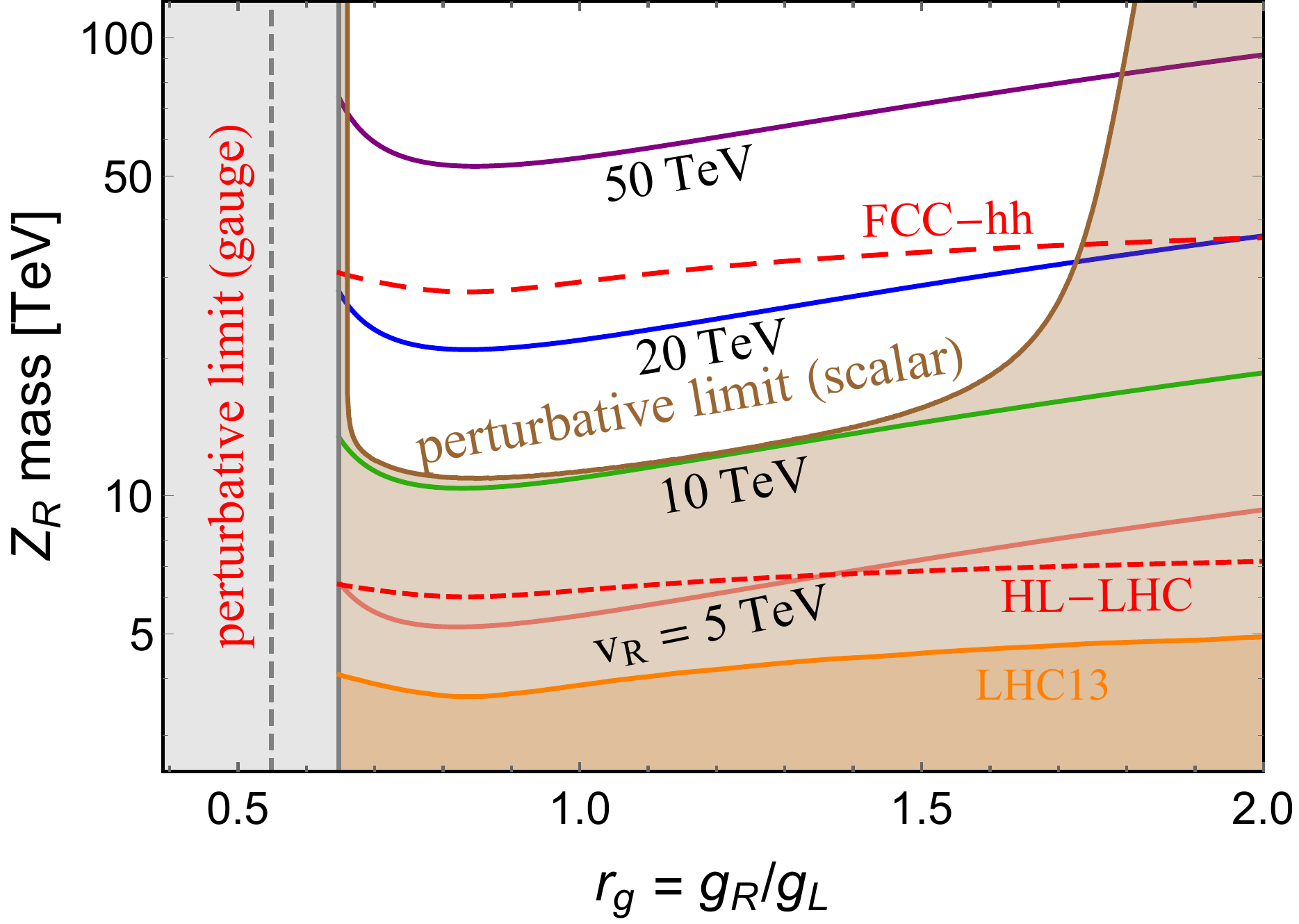}
  \caption{Current LHC13 constraints on the $W_R$ (left) and $Z_R$ (right) masses in the minimal LRSM (shaded orange) as function of $r_g = g_R/g_L$, and future prospects at the HL-LHC 14 TeV with an integrated luminosity of 3000 fb$^{-1}$ (short-dashed red) and the 100 TeV collider FCC-hh with a luminosity of 30 ab$^{-1}$ (long-dashed red). The pink, green, blue and purple lines are the $W_R/Z_R$ mass with the RH scale $v_R = 5$, 10, 20, 50 TeV, respectively. The shaded gray and brown regions are excluded respectively by the perturbative constraints from the gauge and scalar  sectors up to the GUT scale. The dashed vertical line corresponds to the absolute theoretical bound in Eq.~(\ref{eqn:bound}).}
  \label{fig:LRSM3}
\end{figure}

In the LRSM, the couplings of $Z_R$ boson to the SM fermions and the heavy RHNs are the same as in the $U(1)_{B-L}$ model in Section~\ref{sec:B-L}. Thus, the dilepton limits from current LHC 13 TeV data~\cite{ATLAS:2016cyf, CMS:2016abv} and the prospects at the HL-LHC~\cite{Diener:2010sy,Godfrey:2013eta} and future 100 TeV colliders~\cite{Godfrey:2013eta, Rizzo:2014xma} are also the same as in $U(1)_{B-L}$ model, up to the different perturbative windows for the gauge couplings in Eqs.~(\ref{eqn:window2}) and (\ref{eqn:window3}), respectively. The current LHC 13 TeV dilepton constraints on the $Z_R$ mass in the minimal LRSM and the future prospects are shown in the right panel of Fig.~\ref{fig:LRSM3}, along with the contours for $M_{Z_R} (r_g)$ with the RH scale $v_R = 5$, 10, 20 and 50 TeV. In the plot we have also shown the absolute theoretical lower bound on $r_g > \tan\theta_w$ from Eq.~(\ref{eqn:bound}) as the dashed vertical line, which is weaker than the ``real'' lower bound from perturbativity up to the GUT scale shown in Figs.~\ref{fig:LRSM1} and \ref{fig:LRSM2} (the solid vertical gray line in Fig.~\ref{fig:LRSM3}). The scalar perturbativity limit shown in Fig.~\ref{fig:LRSM3} will be discussed in Section~\ref{sec:scalar}.

As for the limits on $W_R$ boson in the LRSM, due to the Majorana nature of the heavy RHNs, the same-sign dilepton plus jets $pp \to W_R \to \ell^\pm N \to \ell^\pm \ell^\pm jj$ is the ``smoking-gun'' signal from the production and decay of the heavy $W_R$ boson at hadron colliders~\cite{Keung:1983uu}. The associated searches of $W_R$ and RHN have been performed at LHC 13 TeV~\cite{Sirunyan:2018pom, Aaboud:2018spl}. To be concrete, we fix the RHN mass $M_N = 1$ TeV; for such a benchmark scenario, the current LHC data requires that the $W_R$ mass $M_{W_R} > 4.7$ TeV for $g_R = g_L$~\cite{Aaboud:2018spl}. If $g_R \neq g_L$, we have to re-evaluate the dependence of the production of $W_R$ and the subsequent decays $W_R \to \ell^\pm N$ and $N \to \ell^\pm \ell\ell$ on the gauge coupling $g_R$.\footnote{The $W_R$ boson might also decay into $WZ$ and $Wh$, with the branching fractions depending largely on the VEV $\kappa'/\kappa$~\cite{Dev:2015pga}. This does not affect the dependence of $W_R$ production on the gauge coupling $g_R$. For simplicity, we have also neglected the effect of the heavy-light neutrino mixing on the $W_R$ decay~\cite{Chen:2013fna}, since this mixing is severely constrained for TeV-scale LRSM with type-I seesaw~\cite{Deppisch:2015qwa}.}  Specifically,
\begin{itemize}
  \item The production of $W_R$ at hadron colliders is proportional to the $W_R$ couplings to the SM quarks, i.e. $\sigma (pp \to W_R) \propto g_R^2$.
  \item The $W_R$ boson decays predominately into the SM quarks and the charged leptons and heavy RHNs, i.e. $W_R \to \bar{q}_R q_R^\prime,\, \ell_R N$. All the partial widths are proportional to $g_R^2$, but not the branching fraction ${\rm BR} (W_R \to \ell N)$.
  \item In the limit of vanishing $W - W_R$ mixing and heavy-light neutrino mixing, $N \to \ell jj$ is the dominant decay mode (assuming $N$ here is the lightest RHN), whose branching fraction does not depend on $g_R$.
\end{itemize}
In short, the $g_R$ dependence is only relevant to the production $pp \to W_R$. For fixed $W_R$ mass, we need only to rescale the production cross section $\sigma (pp \to W_R)$ by a factor of $r_g^2 = (g_R/g_L)^2$. The current LHC constraint on $W_R$ mass is presented in the left panel of Fig.~\ref{fig:LRSM3} as function of $r_g$, along with the contours of $M_{W_R}$ for $v_R = 5$, 10, 20 and 50 TeV.

It is a good approximation in the minimal LRSM that the right-handed quark mixing matrix is identical to the CKM matrix in the SM, up to some unambiguous signs~\cite{Senjanovic:2014pva}. Then the $W_R$-mediated right-handed currents contributes to the $K_0 - \overline{K}_0$ and $B - \overline{B}$ mixings, leading to strong constraints on the $W_R$ mass, $M_{W_R} \gtrsim 3$ TeV~\cite{Beall:1981ze, Mohapatra:1983ae, Ecker:1985vv, Zhang:2007da}.  This limit does not depend on the coupling $g_R$, as in the limit of $m_{K,\,B} \ll M_{W_R}$ the $g_R$ dependence of $W_R$ coupling to the SM quarks is canceled out by the dependence of $g_R$ in the $W$ boson propagator. The $W_R$ contribution is effectively suppressed by $v_{\rm EW}^2/v_R^2$. As the quark flavor limits on $W_R$ mass is significantly lower than that from the direct searches at the LHC for $r_g \gtrsim 0.65$, they are not shown in the left panel of Fig.~\ref{fig:LRSM3}. The $W_R$ contributes also to neutrinoless double $\beta$-decays~\cite{Dev:2013vxa, Dev:2014xea, Mohapatra:1981pm, Hirsch:1996qw, Tello:2010am, Chakrabortty:2012mh, Barry:2013xxa, Huang:2013kma, Bambhaniya:2015ipg}, which however depends on the masses of heavy RHNs and the doubly-charged scalars, and therefore, not included in Fig.~\ref{fig:LRSM3}.

The $W_R$ could be probed up to 5.4 TeV at LHC 14 with a luminosity of 300 fb$^{-1}$~\cite{Ferrari:2000sp, Nemevsek:2018bbt}. By rescaling the production cross section $\sigma (pp \to W_R)$ using {\tt CalcHEP}~\cite{Belyaev:2012qa}, the $W_R$ prospects could go up to 6.5 TeV for $g_R = g_L$ at the HL-LHC where the integrated luminosity is 10 times larger (3000 fb$^{-1}$). At future 100 TeV hadron colliders, for a relatively light RHN $M_N \ll M_{W_R}$, the decay products from the RHN tend to be highly-collimated and form fat jets. We adopt the analysis in Ref.~\cite{Mitra:2016kov} where $M_N/M_{W_R}$ was taken to be $0.1$. Given a luminosity of 30 ab$^{-1}$ at 100 TeV hadron colliders, the $W_R$ mass could be probed up to 38.4 TeV with $g_R = g_L$. 
The projected sensitivity of $W_R$ mass for a relatively low $M_N$ at the HL-LHC and future 100 TeV collider FCC-hh could also be generalized to the case with $g_R \neq g_L$, which are depicted in the left panel of Fig.~\ref{fig:LRSM3} respectively as the short-dashed and long-dashed red curves.

With the heavy gauge boson masses in the minimal LRSM
\begin{eqnarray}
M_{W_R}^2 \ \simeq \ g_R^2 v_R^2 \,, \quad
M_{Z_R}^2 \ \simeq \ 2 (g_R^2 + g_{BL}^2) v_R^2 \,,
\end{eqnarray}
the current direct search limits of the $W_R$ and $Z_R$ boson at LHC 13 TeV and the future prospects at the HL-LHC and FCC-hh can be translated into limits on the $v_R$ scale (as in the $U(1)_{B-L}$ model in Section~\ref{sec:B-L}), which are presented in Fig.~\ref{fig:LRSM4}. For illustration purpose, we have also shown the contours of $M_{W_R} (M_{Z_R}) = 5$, 10, 20 and 50 TeV in the left (right) panel of Fig.~\ref{fig:LRSM4}, which are depicted respectively in pink, green, blue and purple. We find that the RH scale could be probed up to $\simeq 87$ TeV in the searches of $W_R$ boson and $\sim 25$ TeV in the $Z_R$ boson channel at the 100 TeV collider.

\subsection{Perturbativity constraints from the scalar sector}
\label{sec:scalar}

The most general renormalizable scalar potential for the
$\Phi$ and $\Delta_{R}$ fields is given by
\begin{eqnarray}
\label{eqn:potential}
\mathcal{V} & \ = \ & - \mu_1^2 \: {\rm Tr} (\Phi^{\dag} \Phi) - \mu_2^2
\left[ {\rm Tr} (\tilde{\Phi} \Phi^{\dag}) + {\rm Tr} (\tilde{\Phi}^{\dag} \Phi) \right]
- \mu_3^2 \:  {\rm Tr} (\Delta_R
\Delta_R^{\dag}) \nonumber
\\
&&+ \lambda_1 \left[ {\rm Tr} (\Phi^{\dag} \Phi) \right]^2 + \lambda_2 \left\{ \left[
{\rm Tr} (\tilde{\Phi} \Phi^{\dag}) \right]^2 + \left[ {\rm Tr}
(\tilde{\Phi}^{\dag} \Phi) \right]^2 \right\} \nonumber \\
&&+ \lambda_3 \: {\rm Tr} (\tilde{\Phi} \Phi^{\dag}) {\rm Tr} (\tilde{\Phi}^{\dag} \Phi) +
\lambda_4 \: {\rm Tr} (\Phi^{\dag} \Phi) \left[ {\rm Tr} (\tilde{\Phi} \Phi^{\dag}) + {\rm Tr}
(\tilde{\Phi}^{\dag} \Phi) \right]  \\
&& + \rho_1  \left[ {\rm
Tr} (\Delta_R \Delta_R^{\dag}) \right]^2 
+ \rho_2 \: {\rm Tr} (\Delta_R
\Delta_R) {\rm Tr} (\Delta_R^{\dag} \Delta_R^{\dag}) \nonumber
\\
&&+ \alpha_1 \: {\rm Tr} (\Phi^{\dag} \Phi) {\rm Tr} (\Delta_R \Delta_R^{\dag})
+ \left[  \alpha_2 e^{i \delta_2}  {\rm Tr} (\tilde{\Phi}^{\dag} \Phi) {\rm Tr} (\Delta_R
\Delta_R^{\dag}) + {\rm H.c.} \right]
+ \alpha_3 \: {\rm
Tr}(\Phi^{\dag} \Phi \Delta_R \Delta_R^{\dag}) \,.  \nonumber
\end{eqnarray}
Due to the LR symmetry, all the 12 parameters $\mu^2_{1,2,3}$, $\lambda_{1,2,3,4}$, $\rho_{1,2}$, $\alpha_{1,2,3}$ are real, and the only CP-violating phase is $\delta_2$ associated with the coupling $\alpha_2$, as explicitly shown in Eq.~\eqref{eqn:potential}.\footnote{This potential stems from the full LRSM at a higher energy scale in presence of the left-handed triplet $\Delta_L$. At the high scale, all but one of the couplings are real. At low scales there will be small phases in some couplings induced by radiative renormalization group effects. We ignore these small phases. Our main conclusions are not affected by this.} The neutral component of the triplet develops a non-vanishing vacuum expectation value (VEV) $\langle \Delta_R^0 \rangle = v_R$, which breaks the $SU(2)_R \times U(1)_{B-L}$ down to the SM $U(1)_Y$, and generates  masses for the heavy scalars, the $W_R$ and $Z_R$ bosons and the RHNs. The bidoublet VEVs $\langle \phi_1^0 \rangle = \kappa$ and $\langle \phi_2^0 \rangle = \kappa'$ are responsible for the electroweak symmetry breaking. Neglecting the CP violation and up to the leading order in the small parameters $\epsilon = v_{\rm EW} /v_R$ and $\xi = \kappa'/\kappa$, the physical scalar masses are respectively~\cite{Dev:2016dja}
\begin{align}
\label{eqn:masses}
M_h^2 & \ \simeq \ 4 \lambda_1 v_{\rm EW}^2 \,, &
M^2_{H_1,\, A_1,\, H_1^\pm} & \ \simeq \ \alpha_3 v_R^2 \,, \nonumber \\
M_{H_3}^2 & \ \simeq \ 4 \rho_1 v_R^2 \,, &
M_{H_2^{\pm\pm}}^2 & \ \simeq \ 4 \rho_2 v_R^2 \,,
\end{align}
where $h$ is the SM Higgs, $H_1$, $A_1$ and $H_1^\pm$ respectively the heavy CP-even and CP-odd neutral components and the singly-charged scalars from the bidoublet $\Phi$, $H_3$ and $H_2^{\pm\pm}$ are the neutral and doubly-charged scalars from the triplet $\Delta_R$, following the convention of Ref.~\cite{Dev:2016dja}. 

In the minimal LRSM, the heavy neutral scalars $H_1$ and $A_1$ have tree-level FCNC couplings to the SM quarks, which contribute to the $K_0 - \overline{K}_0$, $B_d - \overline{B}_d$ and $B_s - \overline{B}_d$ mixings. Thus their masses are tightly constrained by the high-precision flavor data, i.e. $M_H \gtrsim 10$ TeV~\cite{Ecker:1983uh, Zhang:2007da, Maiezza:2010ic}. For a few-TeV scale $v_R$, this implies that the quartic coupling $\alpha_3 \simeq M_{H_1}^2 / v_R^2$ is pretty large, typically of order one. The RG running of the quartic couplings in Eq.~(\ref{eqn:potential}) are all entangled together, and a large $\alpha_3$ is the main reason why the LRSM could easily hit a Landau pole at an energy scale that is much lower than the GUT scale~\cite{Rothstein:1990qx, Chakrabortty:2013zja, Chakrabortty:2016wkl, Maiezza:2016ybz}. This could be alleviated if the $v_R$ scale is higher and the coupling $\alpha_3$ gets smaller. Therefore, the perturbativity of the quartic couplings in Eq.~(\ref{eqn:potential}) up to the GUT scale could set a lower bound on the $v_R$ scale, assuming there is no intermediate scales or particles in between $v_R$ and the GUT scale.

A thorough analysis of the RG running of all the quartic couplings in Eq.~(\ref{eqn:potential}) is rather complicated and it obfuscates the perturbativity limits on the $v_R$ scale. While some of the quartic couplings could be tuned very small at the $v_R$ scale as they only induce mixings among the scalars such as $\alpha_{1,\,2}$~\cite{Dev:2016dja}, there are only four quartic couplings, i.e. the $\lambda_1$, $\alpha_3$, $\rho_1$ and $\rho_2$ appearing in Eq.~(\ref{eqn:masses}), that are responsible for the scalar masses at the tree level. Therefore, for the purpose of perturbativity limits in the scalar sector, we consider a simple scenario with only these four non-vanishing quartic couplings $\lambda_1$, $\alpha_3$, $\rho_1$ and $\rho_2$ at the $v_R$ scale. In particular, we set the scalar masses to the following benchmark values:
\begin{align}
\label{eqn:masses2}
M_h & \ = \ 125 \, {\rm GeV} \,, &
M_{H_1,\, A_1,\, H_1^\pm} & \ = \ 10 \, {\rm TeV} \,, \nonumber \\
M_{H_3} & \ = \ 100 \, {\rm GeV} \,, &
M_{H_2^{\pm\pm}} & \ = \ 1 \, {\rm TeV} \,,
\end{align}
from which one could obtain the values of $\lambda_1$, $\alpha_3$, $\rho_1$ and $\rho_2$ by using Eq.~(\ref{eqn:masses}). All other quartic couplings $\lambda_{2,\,3,\,4}$, $\alpha_{1,2}$ are set to zero, and this corresponds to the limits without any tree-level scalar mixing at the $v_R$ scale. In the limit of vanishing mixing between $h$ and $H_1$, the neutral scalar $H_3$ from the triplet $\Delta_R$ is hadrophobic and the experimental constraints on $H_3$ are rather weak~\cite{Dev:2016vle, Dev:2017dui}. Thus we have set $H_3$ to be light, at the 100 GeV scale, in Eq.~(\ref{eqn:masses2}). The smoking-gun signal of a doubly-charged scalar is the same-sign dilepton pairs $H_2^{\pm\pm} \to \ell_\alpha^\pm \ell_\beta^\pm$ with $\alpha,\,\beta = e,\, \mu,\, \tau$, which is almost background free. The current most stringent limits are from the LHC 13 TeV data~\cite{Aaboud:2017qph, CMS:2017pet}, which requires that $M_{H_2^{\pm\pm}} \gtrsim (271 - 760)$ GeV, depending largely on the charged lepton flavors involved~\cite{Dev:2018kpa}. To be concrete, we have set the doubly-charged scalar mass at 1 TeV in Eq.~(\ref{eqn:masses2}), which easily satisfies the current LHC constraints. As for the bidoublet masses $M_{H_1,\, A_1,\, H_1^\pm}$, we have taken the minimum possible value allowed by FCNC constraints~\cite{Zhang:2007da}, whereas for the SM Higgs, we have taken the current best-fit value~\cite{Aad:2015zhl}.

All the RGEs for the gauge couplings $g_{L,\,R,\,BL}$ and the quartic couplings in the potential in Eq.~(\ref{eqn:potential}) are collected in Appendix~\ref{sec:RGE:LRSM} up to the two-loop level. To be self-consistent, we include also the RGE for the dominant Yukawa coupling $h_t$ that is responsible for generation of the top quark mass at the electroweak scale. The Yukawa couplings for the bottom and tauon are comparatively much smaller and are neglected here. For a RH scale $v_R \gtrsim 10$ TeV, the Yukawa coupling $f_R$ of $\Delta_R$ to the lepton doublets are also small if the masses of the three RHNs $M_N \simeq $ TeV. For simplicity, the $f_R$ terms in the $\beta$-functions are also neglected. See Appendix~\ref{sec:RGE:LRSM} for more details.

\begin{figure}[!t]
  \centering
  \includegraphics[width=0.495\linewidth]{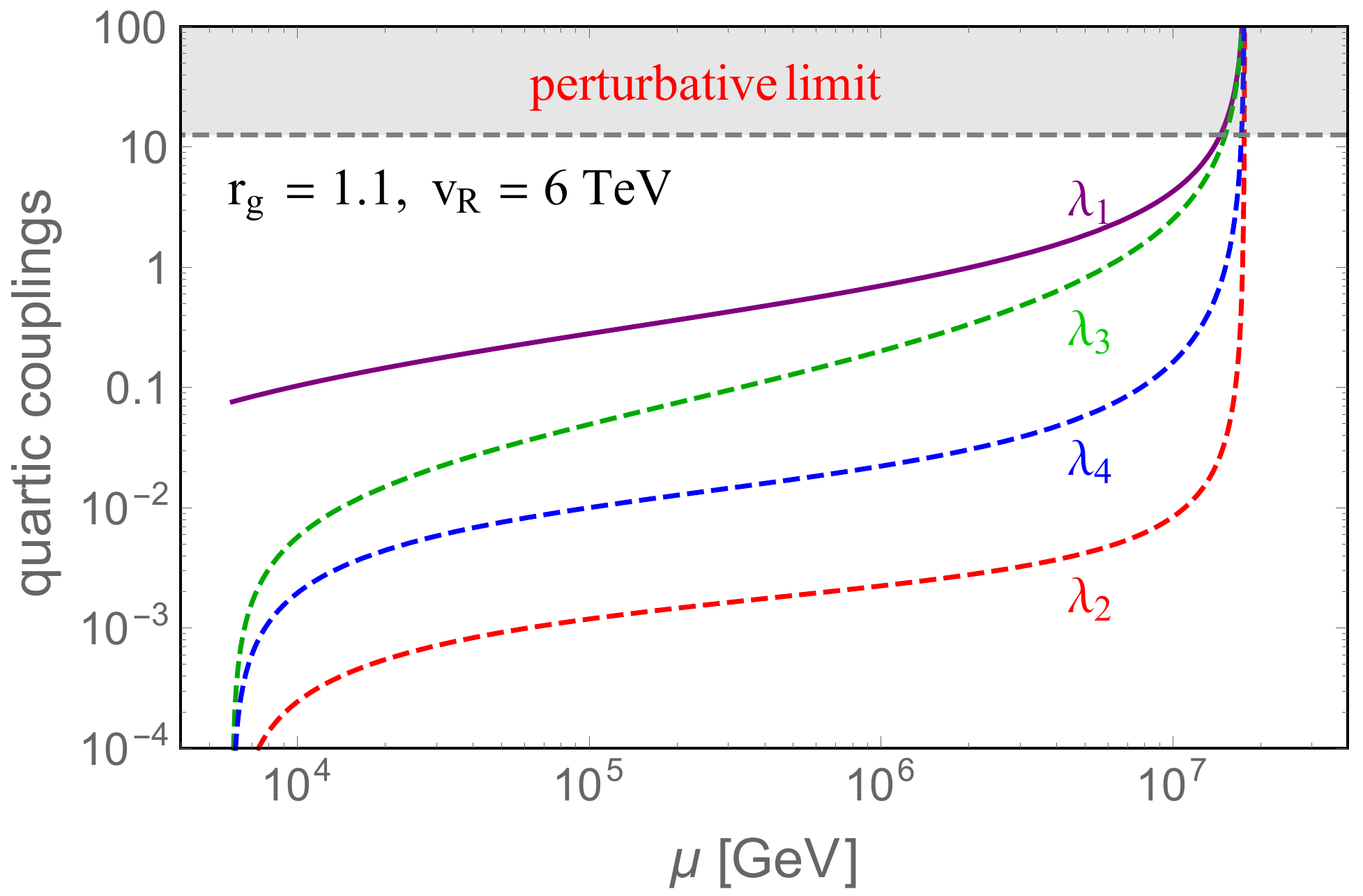}
  \includegraphics[width=0.482\linewidth]{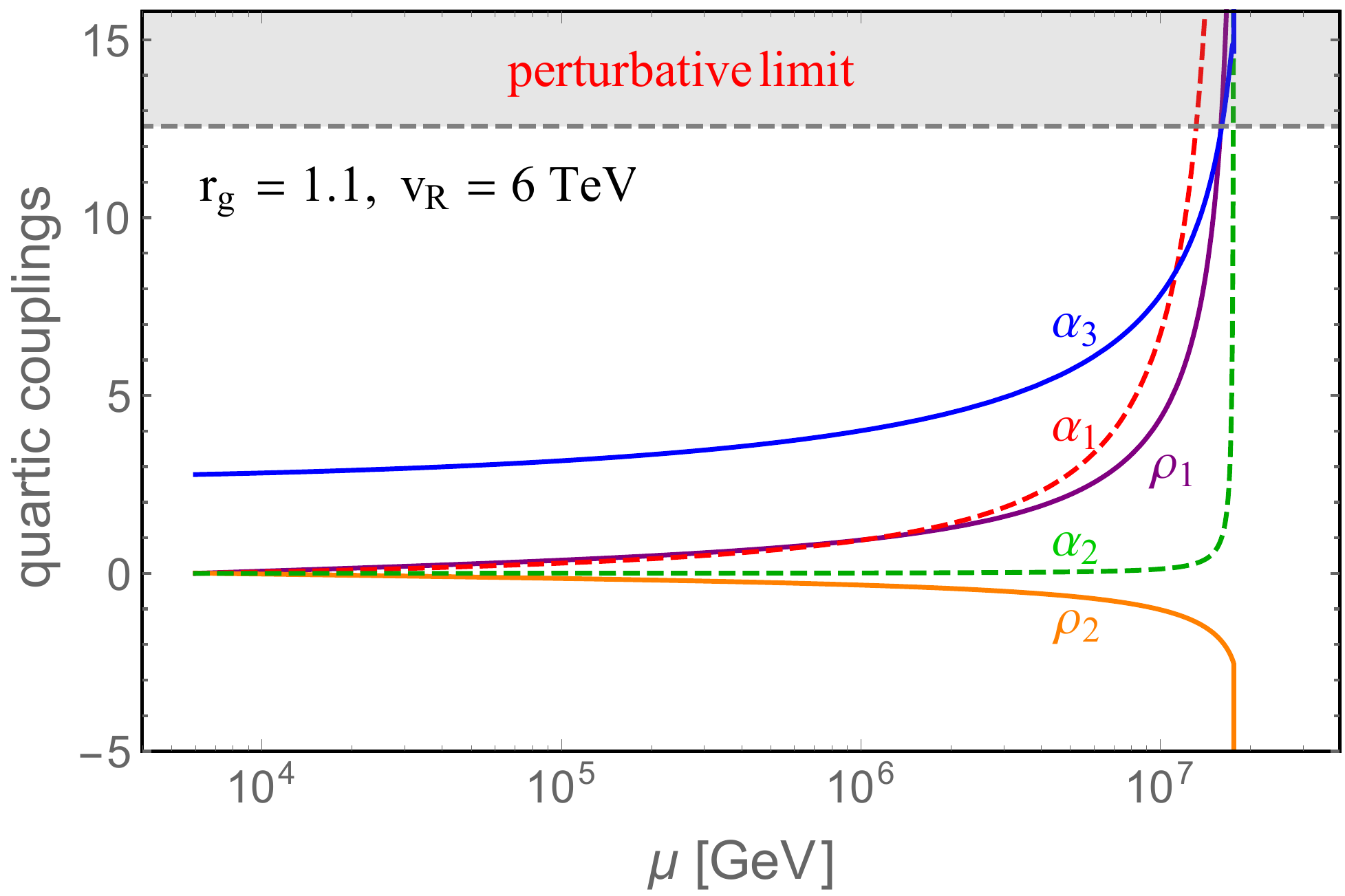} \vspace{5pt} \\
  \includegraphics[width=0.49\linewidth]{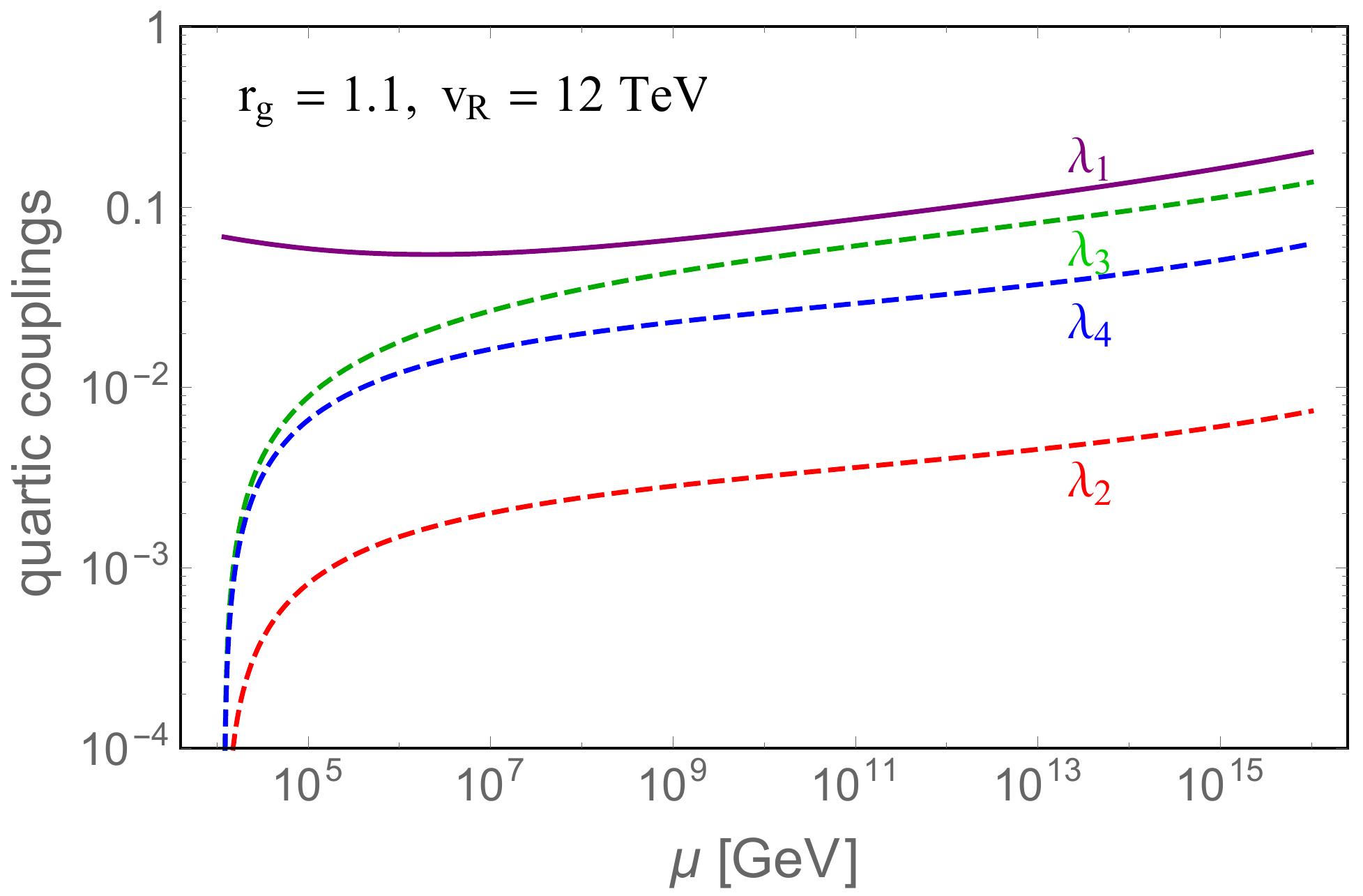}
  \includegraphics[width=0.495\linewidth]{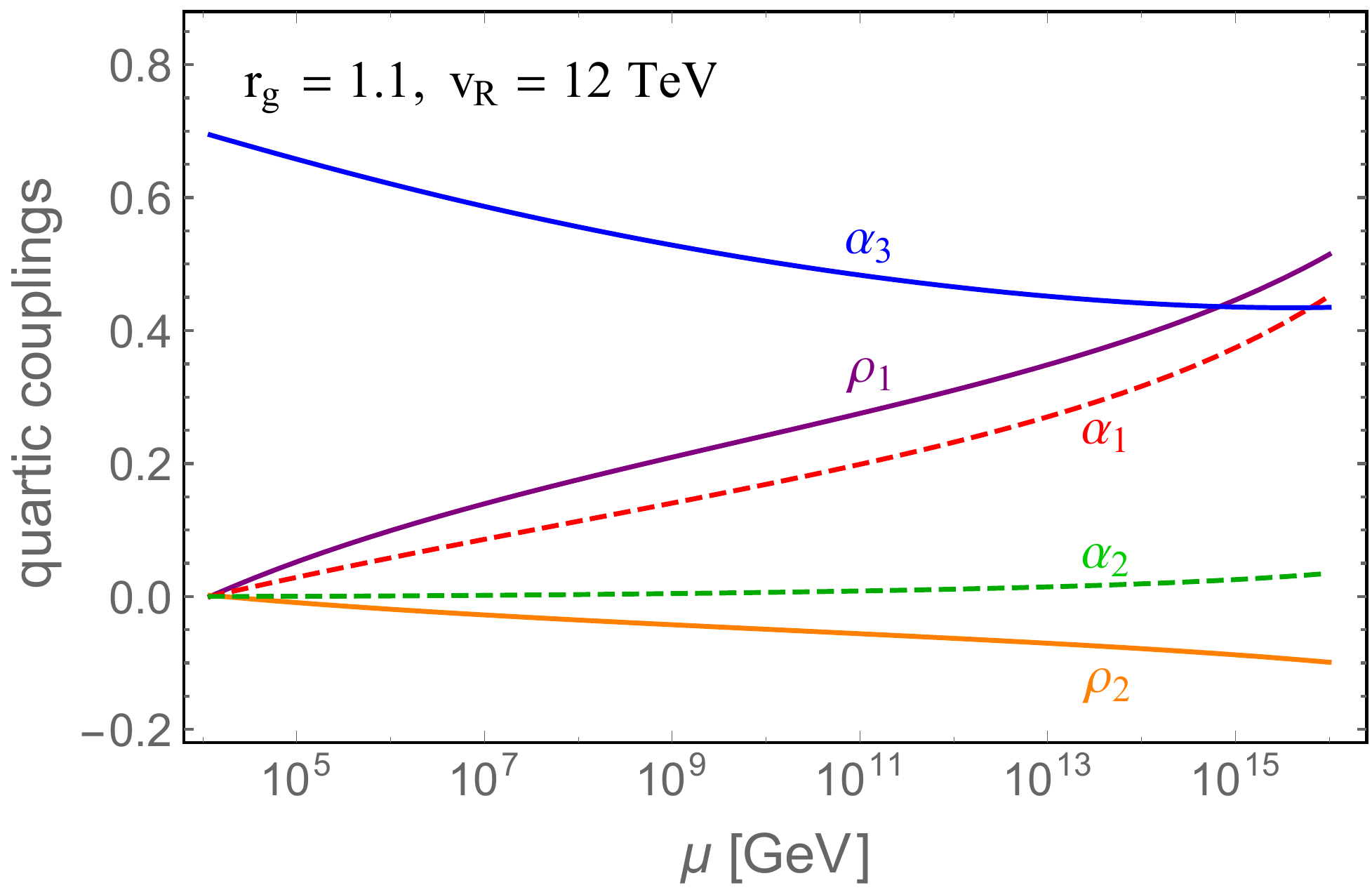}
  \caption{RG running of the quartic couplings $\lambda_{1,\, 2,\, 3,\, 4}$ (left), and $\rho_{1,\,2}$, $\alpha_{1,\,2,\,3}$ (right) in the scalar potential Eq.~(\ref{eqn:potential}) of minimal LRSM from $v_R$ up to the GUT scale, with $r_g = 1.1$ and $v_R = 6$ TeV (upper panels),  $v_R = 12$ TeV (lower panels). }
  \label{fig:quartic}
\end{figure}

Given the scalar masses in Eq.~(\ref{eqn:masses2}), all the $\beta$-functions for the quartic couplings in Eq.~(\ref{eqn:LR:beta1}) to (\ref{eqn:LR:beta9}) are dominated by the $\alpha_3$ terms if the RH scale $v_R$ is not too much higher than the TeV scale, i.e.
\begin{eqnarray}
\label{eqn:beta:example}
16\pi^2 \beta (\lambda_1) & \ = \ &
\frac54 \alpha_3^2 +
\frac38 \left( 3 g_L^4 + 2 g_L^2 g_R^2 + 3 g_R^4 \right)
- 6 h_t^4 + \cdots \,,
\end{eqnarray}
where the dots stand for the subleading terms. For a few-TeV $v_R$ and
$\alpha_3 \gtrsim {\cal O} (1)$, the quartic couplings rapidly blow up before reaching the GUT scale~\cite{Rothstein:1990qx, Chakrabortty:2013zja, Chakrabortty:2016wkl, Maiezza:2016ybz}. An explicit example is shown in the two upper panels of Fig.~\ref{fig:quartic}, with $r_g = g_R/g_L = 1.1$ and $v_R = 6$ TeV, where the quartic couplings become non-perturbative at $\sim 10^7$ GeV. When the RH scale $v_R$ is higher, for a fixed mass $M_{H_1} = 10$ TeV, the coupling $\alpha_3 \simeq M_{H_1}^2 / v_R^2$ is significantly smaller. As a result, in a large region of the parameter space, all the quartic couplings are perturbative up to the GUT scale, as exemplified in the two lower panels of  Fig.~\ref{fig:quartic} with $r_g = g_R/g_L = 1.1$ and $v_R = 12$ TeV. In both examples, the bounded-from-below conditions in the scalar sector are respected~\cite{Chakrabortty:2016wkl, Dev:2018foq}:\footnote{More generic vacuum stability criteria can be found, e.g., in Ref.~\cite{Chakrabortty:2013mha}.}
\begin{eqnarray}
\lambda_1 \geq 0 \,, \quad
\rho_1 \geq 0 \,, \quad
\rho_1 + \rho_2 \geq 0 \,, \quad
\rho_1 + 2\rho_2 \geq 0 \,.
\end{eqnarray}

One should note that the $g_R$ and $g_{BL}$ terms in the $\beta$-functions in Eqs.~(\ref{eqn:LR:beta1}) to (\ref{eqn:LR:beta9}) might be unacceptably large. Thus the perturbativity limits in the scalar sector depend also on the gauge couplings $g_R$ and $g_{BL}$, or equivalently the ratio $r_g = g_R/g_L$. As seen in Eq.~(\ref{eqn:beta:example}), when $g_R \gtrsim {\cal O} (1)$ [or $g_{BL} \gtrsim {\cal O} (1)$], the constraints on the $v_R$ scale and $\alpha_3$ tend to be more stringent. The $r_g$-dependent scalar perturbativity constraints on $v_R$ are shown in Fig.~\ref{fig:LRSM4} as the shaded brown regions. Numerically, we find the requirement in the minimal LRSM that
\begin{eqnarray}
v_R \gtrsim 10 \,\, {\rm TeV} \quad \text{for} \quad
0.65 \lesssim r_g \lesssim 1.6 \,,
\end{eqnarray}
which makes the perturbativity constraints very stringent in the gauge sector (see Figs.~\ref{fig:LRSM1} and \ref{fig:LRSM2}). The quartic couplings blow up very quickly when $r_g$ is out of this range unless $v_R$ is much higher than 10 TeV, as $g_R \gtrsim {\cal O} (1)$ or $g_{BL} \gtrsim {\cal O} (1)$.

It is remarkable that the perturbativity constraints from the scalar sector supersede the current LHC constraints on the $W_R$ and $Z_R$ bosons in the minimal LRSM, and even the projected $Z_R$ sensitivity at the HL-LHC, leaving only a very narrow window for $W_R$, as shown in Fig.~\ref{fig:LRSM3}. Fortunately, future 100 TeV colliders could probe a much larger parameter space. All the numerical ranges of the maximum $W_R$ and $Z_R$ mass reach and the corresponding $v_R$ scales at future hadron colliders are collected in Table~\ref{tab:LRSM_limit}, with both the gauge and scalar perturbativity constraints taken into consideration. Finding a heavy $W_R$ and/or $Z_R$ boson at the HL-LHC, would have strong implications for the interpretation in the minimal LRSM. For instance, if a $Z_R$ boson was to be found at the LHC, then it does not belong to the minimal LRSM. It could still be accommodated in the LRSM framework by introducing some exotic particles or an intermediate scale, e.g. at $\sim 10^{6}$ GeV, to the minimal LRSM to keep all the gauge, Yukawa and quartic couplings perturbative up to the GUT scale.

\begin{table}[!t]
  \centering
  \caption{Lower bounds on the $W_R$ and $Z_R$ boson masses and the corresponding $v_R$ scale in the minimal LRSM from the current LHC13 data~\cite{Aaboud:2018spl, ATLAS:2016cyf, CMS:2016abv} and the prospects at the HL-LHC 14 TeV with an integrated luminosity of 3000 fb$^{-1}$~\cite{Ferrari:2000sp, Diener:2010sy,Godfrey:2013eta} and future 100 TeV collider FCC-hh with a luminosity of 30 ab$^{-1}$~\cite{Mitra:2016kov, Godfrey:2013eta, Rizzo:2014xma}, with both the gauge and scalar perturbativity limits up to the GUT scale taken into consideration. The range in each case corresponds to the allowed range of $r_g$ from perturbativity constraints, as given in Figs.~\ref{fig:LRSM3} and \ref{fig:LRSM4}. 
The missing entries mean that the corresponding maximum experimental reach has been excluded by the scalar perturbativity constraints. See text for more details. }
  \label{tab:LRSM_limit}
  \begin{tabular}{l c c c c}
  \hline\hline
  \multirow{2}{*}{collider } &  \multicolumn{2}{c}{$W_R$ searches} & \multicolumn{2}{c}{$Z_R$ searches} \\ \cline{2-5}
  & $M_{W_R}$ [TeV] & $v_R$ [TeV] &  $M_{Z_R}$ [TeV] & $v_R$ [TeV]  \\ \hline
  LHC13 & $-$ & $-$ & $-$ & $-$  \\
  HL-LHC & $[6.09,\, 6.47]$ & $[10.3,\, 14.8]$ & $-$ & $-$ \\
  FCC-hh & $[35.6, 42.2]$ & $[38.3,\, 87.5]$ & $[27.9,\, 35.4]$ & $[21.8,\, 26.8]$ \\ \hline

  \end{tabular}
\end{table}

It should be emphasized that the perturbative constraints on the $W_R$ and $Z_R$ masses and the $v_R$ scale from the scalar sector shown in Figs.~\ref{fig:LRSM3} and \ref{fig:LRSM4} are based on the assumptions of the scalar masses in Eq.~(\ref{eqn:masses2}) and the vanishing quartic couplings $\lambda_{2,\,3,\,4}$, $\alpha_{1,2}$. In light of the stringent flavor constraints on the bidoublet scalars $H_1$, $A_1$ and $H_1^\pm$, the dilepton constraints on the doubly-charged scalar $H_2^{\pm\pm}$ and the flavor constraints on the neutral scalar $H_3$, the masses in Eq.~(\ref{eqn:masses2}) are almost the most optimistic values allowed by the current data that could give us the most conservative perturbativity constraints. If the BSM Higgs masses get larger (especially in the bidoublet sector), the corresponding quartic couplings $\alpha_3$, $\rho_1$ and $\rho_2$ will be accordingly enhanced by $M^2 / v_R^2$ at the $v_R$ scale (with $M$ standing for the generic BSM scalar mass), and the scalar perturbativity limits on the $v_R$ scale would be more stringent. Furthermore, if the couplings $\lambda_{2,\,3,\,4}$, $\alpha_{1,2}$ are not zero at the $v_R$ scale, the quartic couplings tend to hit the Landau pole at a lower scale. 

We have also checked also the two-loop corrections to the scalar perturbativity limits by deriving all the two-loop RGEs using the code {\tt PyR@TE}~\cite{Lyonnet:2013dna, Lyonnet:2015jca}, as collected in Eqs.~(\ref{eqn:LR:gauge:beta1}) to (\ref{eqn:LR:beta10}). It turns out that the two-loop corrections only amount to less than 3\% different for the scalar perturbativity limits on $v_R$ and the heavy gauge boson masses, as compared to the one-loop results presented in this section.

In the limit of small scalar mixing, just as we have assumed above, $\lambda_1$ can be identified as the SM quartic coupling. As a byproduct, the extra scalars in the LRSM contribute positively to the $\beta (\lambda_1)$ in Eq.~(\ref{eqn:LR:beta1}) in a larger region of parameter space, which helps to stabilize the SM vacuum up to the GUT scale or even up to the Planck scale. The full analysis of the stability of  the scalar potential is beyond the main scope of this paper. See Ref.~\cite{Dev:2018foq} for a recent analysis in this direction.

\section{Conclusion}
\label{sec:conclusion}

In conclusion, we find that in the extensions of the electroweak gauge group to either $SU(2)_L\times U(1)_{I_{3R}}\times U(1)_{B-L}$ or the left-right symmetric group $SU(2)_L \times SU(2)_R \times U(1)_{B-L}$, both of which contribute to the electric charge, there are strong limits on the new gauge couplings $g_R$ and $g_{BL}$ from the requirement that the couplings remain perturbative till the GUT scale. We obtain those limits for the minimal versions of these models and study their implications for collider phenomenology. We find in particular that the ratio $r_g\equiv g_R/g_L$, or effectively the gauge couplings $g_R$ and $g_{BL}$, are limited to a very narrow range at the TeV scale, as seen in Figs.~\ref{fig:UoneBL_g2} and \ref{fig:LRSM2}. Inclusion of the scalar sector in the minimal LRSM implies that the RH symmetry breaking scale in LRSM must have a lower bound of about 10 TeV for a limited coupling range $0.65\lesssim r_g \lesssim 1.6$. The gauge (and scalar) perturbative constraints have rich implications for the searches of $Z_R$ (and $W_R$) bosons in these models at the HL-LHC and future 100 TeV colliders. All the direct search constraints on the $W_R$ and $Z_R$ masses from LHC 13 TeV, as well as the future prospects at HL-LHC and 100 TeV colliders, depend on the BSM gauge couplings $g_R$ and $g_{BL}$ (or effectively the ratio $r_g$). All the $Z_R$ (and $W_R$) mass ranges and the corresponding $v_R$ scales are collected in Figs.~\ref{fig:UoneBL_ZR}, \ref{fig:UoneBL_vR} and Table~\ref{tab:UoneBL_limit} for the $U(1)_{B-L}$ model, and Figs.~\ref{fig:LRSM3}, \ref{fig:LRSM4} and Table~\ref{tab:LRSM_limit} for the LRSM. One of the most striking results we find is that the perturbativity constraints already exclude the possibility of finding the $Z_R$ boson belonging to the minimal LRSM at the HL-LHC, and leave only a narrow window for the $W_R$ boson. We hope this serves as an additional motivation for the 100 TeV collider, where a much broader parameter space can be probed.

\section*{Acknowledgments}
We thank K. S. Babu and Florian Staub for useful discussions. The work of B.D.\ is supported by the US Department of Energy under Grant No.\ DE-SC0017987. The work of R.N.M.\ is supported by the US National Science Foundation under Grant No.\ PHY1620074. Y.Z.\ would like to thank the HKUST Jockey Club Institute for Advanced Study, Hong Kong University of Science and Technology and the Center for High Energy Physics, Peking University for generous hospitality where the paper was revised.

\appendix

\section{Two-loop RGEs for the minimal LRSM}
\label{sec:RGE:LRSM}

Here we list the $\beta$-functions for the gauge couplings $g_{S,\,L,\, R,\, BL}$, the quartic couplings $\lambda_{1,\, 2,\, 3,\, 4}$, $\rho_{1,\,2}$ and $\alpha_{1,\,2,\,3}$ in the scalar potential (\ref{eqn:potential}) and the Yukawa coupling $h_t$ in the minimal LRSM up to two-loop level, which are obtained by using {\tt PyR@TE}~\cite{Lyonnet:2013dna, Lyonnet:2015jca}:\footnote{Note that some of the one-loop coefficients obtained here are different from those in Refs.~\cite{Rothstein:1990qx, Chakrabortty:2013zja}. In particular, the coefficient for $g_R$ in these references is $-5/2$, while, with the same matter and scalar fields that contribute to the running of $g_R$, we found it is $-7/3$.}

\begin{eqnarray}
\label{eqn:LR:gauge:beta1}
\beta (g_S) & \ = \ &
\frac{1}{16\pi^2}
\left[ -7 \, g_S^3 \right]
+ \frac{1}{(16\pi^2)^2}
\bigg[ \frac{1}{6} g_S^3 \left(2 g_{BL}^2+3 \left(9 g_L^2+9 g_R^2-52 g_S^2-8 h_t^2\right)\right) \bigg] \,, \\
\label{eqn:LR:gauge:beta2}
\beta (g_L) & \ = \ &
\frac{1}{16\pi^2}
\left[ -3 \, g_L^3 \right]
+ \frac{1}{(16\pi^2)^2}
\bigg[ g_L^3 \left(g_{BL}^2+3 \left(g_R^2+4 g_S^2-h_t^2\right)+8 g_L^2\right) \bigg] \,, \\
\label{eqn:LR:gauge:beta3}
\beta (g_R) & \ = \ &
\frac{1}{16\pi^2}
\left[ -\frac73 \, g_R^3 \right]
+ \frac{1}{(16\pi^2)^2}
\bigg[ \frac{1}{3} g_R^3 \left(27 g_{BL}^2+9 g_L^2+80 g_R^2+36 g_S^2-9 h_t^2\right) \bigg] \,, \\
\label{eqn:LR:gauge:beta4}
\beta (g_{BL}) & \ = \ &
\frac{1}{16\pi^2}
\left[ \frac{11}{3} \, g_{BL}^3 \right]
+ \frac{1}{(16\pi^2)^2}
\bigg[ \frac{1}{9} g_{BL}^3 \left(122 g_{BL}^2+3 \left(9 g_L^2+81 g_R^2+8g_S^2-2 h_t^2\right)\right) \bigg] \,, \\
\label{eqn:LR:beta1}
\beta (\lambda_1) & \ = \ &
\frac{1}{16\pi^2} \left[
\frac38 \left( 3 g_L^4 + 2 g_L^2 g_R^2 + 3 g_R^4 \right)
+ 16 (2\lambda_1^2 + 4\lambda_2^2 + \lambda_3^2 + 3 \lambda_4^2) \right. \nonumber \\
&& \left. +3 \alpha_1^2 + 3 \alpha_1 \alpha_3 + \frac54 \alpha_3^2
- 6 h_t^4
- 9 \lambda_1 \left( g_L^2 + g_R^2 \right)
+ 16 \lambda_1 \lambda_3 + 12 \lambda_1 h_t^2 \right] \nonumber \\
&& + \frac{1}{(16\pi^2)^2} \bigg[
\frac{291 g_L^6}{16}+\frac{235 g_R^6}{16}
+g_L^4 \left(-\frac{191 g_R^2}{16}-\frac{9 h_t^2}{4}+\frac{69 \lambda _1}{8}+15 \lambda
   _3\right) \nonumber \\
   &&  +g_{BL}^2 \left(
   24 \alpha _1^2
   +10 \alpha _3^2
   +24 \alpha _1 \alpha _3
   +\frac{5}{3} \lambda _1
   h_t^2-\frac{2}{3} h_t^4
   \right) \nonumber \\
   && -\frac{1}{48} g_L^2 \left(12 g_R^2 \left(30 h_t^2-69 \lambda _1-56 \lambda _3\right)
   +629 g_R^4 \right. \nonumber \\
   && \left. -72 \left(15 \lambda _1 h_t^2+8
   \left(13 \lambda _1^2+8 \lambda _3 \lambda _1+4 \lambda _3^2+18 \lambda _4^2\right)\right)\right) \nonumber \\
   && +g_R^4 \left(30 \alpha _1+15 \alpha
   _3-\frac{9 h_t^2}{4}+\frac{157 \lambda _1}{8}+15 \lambda _3\right) \nonumber \\
   &&+g_R^2 \left(48 \alpha _1^2+17 \alpha _3^2+48 \alpha _1 \alpha
   _3+\frac{45}{2} \lambda _1 h_t^2+156 \lambda _1^2+48 \lambda _3^2+216 \lambda _4^2+96 \lambda _1 \lambda _3\right) \nonumber \\
   && -30 \alpha _1^2 \lambda _1-24 \alpha _2^2 \lambda _1-\frac{29}{2} \alpha _3^2 \lambda _1-30 \alpha _1 \alpha _3 \lambda _1-192 \alpha _2^2
   \lambda _2-96 \alpha _2^2 \lambda _3-144 \alpha _1 \alpha _2 \lambda _4 \nonumber \\
  && -72 \alpha _2 \alpha _3 \lambda _4-12 \alpha _1^3-\frac{13 \alpha
   _3^3}{2}-48 \alpha _1 \alpha _2^2-19 \alpha _1 \alpha _3^2-18 \alpha _1^2 \alpha _3-24 \alpha _2^2 \alpha _3 \nonumber \\
   && +16 g_S^2 h_t^2 \left( 5 \lambda _1 -2 h_t^2 \right)
      +36 h_t^6
      -12 \lambda _1 h_t^4
   -96 h_t^2 \left(
      2 \lambda _1^2
   +4 \lambda _2^2
   + \lambda _3^2
   +3 \lambda_4^2
   + \lambda _1 \lambda _3  \right)
   \nonumber \\
   && -456 \lambda _1^3
    -384 \lambda _3^3
    -3456 \lambda _1 \lambda_2^2
    -704 \lambda _1 \lambda _3^2
   -2208 \lambda _1 \lambda _4^2
   -3328 \lambda _2 \lambda _4^2 \nonumber \\
   &&
   -1792 \lambda _3 \lambda _4^2
   -352 \lambda _1^2 \lambda _3
   -5632 \lambda _2^2 \lambda _3 \bigg] \,, \\
\label{eqn:LR:beta2}
\beta (\lambda_2) & \ = \ &
\frac{1}{16\pi^2} \left[
12 \lambda_4^2 + 3\alpha_2^2 + \frac{3}{16} h_t^4
- 9 \lambda_2 (g_L^2 + g_R^2)
+ 24 \lambda_2 (\lambda_1 + 2 \lambda_3)
+ 12 \lambda_2 h_t^2 \right] \nonumber \\
&& + \frac{1}{(16\pi^2)^2} \bigg[
-24 \alpha _2^2 \lambda _1-30 \alpha _1^2 \lambda _2+24 \alpha _2^2 \lambda _2+\frac{3}{2} \alpha _3^2 \lambda _2-30 \alpha _1 \alpha _3
   \lambda _2-24 \alpha _2^2 \lambda _3 \nonumber \\
   && -36 \alpha _1 \alpha _2 \lambda _4-18 \alpha _2 \alpha _3 \lambda _4-24 \alpha _1 \alpha _2^2-12 \alpha
   _2^2 \alpha _3+24 \alpha _2^2 g_{BL}^2+\frac{5}{3} \lambda _2 g_{BL}^2 h_t^2 \nonumber \\
   && +g_L^2 \left(\frac{57}{4} \lambda _2
   g_R^2+\frac{45}{2} \lambda _2 h_t^2-\frac{9 h_t^4}{32}+54 \lambda _4^2+72 \lambda _1 \lambda _2+288 \lambda _2 \lambda _3\right) \nonumber \\
   && +g_R^2
   \left(48 \alpha _2^2+\frac{45}{2} \lambda _2 h_t^2-\frac{9 h_t^4}{32}+54 \lambda _4^2+72 \lambda _1 \lambda _2+288 \lambda _2 \lambda
   _3\right) \nonumber \\
   && +80 \lambda _2 g_S^2 h_t^2-\frac{231}{8} \lambda _2 g_L^4-\frac{143}{8} \lambda _2 g_R^4-72 \lambda _4^2 h_t^2-36 \lambda _2    h_t^4 \nonumber \\
   && -144 \lambda _1 \lambda _2 h_t^2-288 \lambda _2 \lambda _3 h_t^2+384 \lambda _2^3-512 \lambda _2 \lambda _3^2-432 \lambda _1 \lambda_4^2 \nonumber \\
   && -1248 \lambda _2 \lambda _4^2-480 \lambda _3 \lambda _4^2-488 \lambda _1^2 \lambda _2-1312 \lambda _1 \lambda _2 \lambda _3 \bigg] \,, \\
\label{eqn:LR:beta3}
\beta (\lambda_ 3) & \ = \ &
\frac{1}{16\pi^2} \bigg[
{\frac {27} {8} h_t^4 +   6 \alpha_ 2^2 - \frac {1} {2}\alpha_ 3^2  + \frac {3} {2} g_L^2 g_R^2 -
  9 \lambda_ 3(g_L^2+g_R^2)}    \nonumber \\
&& { + 12 \lambda_3 h_t^2  +   8 (3 \lambda_ 1\lambda_ 3 +   16 \lambda_ 2^2 + 2 \lambda_ 3^2 + 3 \lambda_ 4^2 )} \bigg] \nonumber \\
&& + \frac{1}{(16\pi^2)^2} \bigg[
\frac{291 g_L^6}{32}-\frac{1}{32} \left(415 g_R^2+6 \left(6 h_t^2-23 \lambda _1-308 \lambda _2+114 \lambda _3\right)\right) g_L^4 \nonumber \\
   &&-\frac{1}{96}
   \left(1525 g_R^4+12 \left(30 h_t^2-165 \lambda _1+228 \lambda _2-298 \lambda _3\right) g_R^2 \right. \nonumber \\
   &&  -72 \left(15 \left(\lambda _1-4 \lambda _2+2
   \lambda _3\right) h_t^2+8 \left(13 \lambda _1^2-4 \left(6 \lambda _2-5 \lambda _3\right) \lambda _1 \right. \right. \nonumber \\
   && \left. \left. \left. +2 \left(72 \lambda _2^2-48 \lambda _3
   \lambda _2+8 \lambda _3^2+9 \lambda _4^2\right)\right)\right)\right) g_L^2 \nonumber \\
   && +\frac{235 g_R^6}{32}+6 h_t^6-6 \alpha _1^3-\frac{5 \alpha
   _3^3}{4}-228 \lambda _1^3-768 \lambda _2^3-192 \lambda _3^3+12 g_{BL}^2 \alpha _1^2 \nonumber \\
   && -24 \alpha _1 \alpha _2^2+g_{BL}^2 \alpha
   _3^2-\frac{11}{2} \alpha _1 \alpha _3^2-96 h_t^2 \lambda _1^2-960 h_t^2 \lambda _2^2 \nonumber \\
   && -4800 \lambda _1 \lambda _2^2-144 h_t^2 \lambda
   _3^2-896 \lambda _1 \lambda _3^2+1024 \lambda _2 \lambda _3^2
   -36 \alpha _2 \alpha _3 \lambda _4 \nonumber \\
   && -144 h_t^2 \lambda _4^2-1104 \lambda _1 \lambda _4^2-1216 \lambda _2 \lambda
   _4^2-1120 \lambda _3 \lambda _4^2-9 \alpha _1^2 \alpha _3 \nonumber \\
   && -12 \alpha _2^2 \alpha _3+12 g_{BL}^2 \alpha _1 \alpha _3-18 h_t^4 \lambda
   _1+\frac{5}{6} g_{BL}^2 h_t^2 \lambda _1
   -72 \alpha _1 \alpha _2   \lambda _4 \nonumber \\
   && +40 g_S^2 h_t^2 \lambda _1-15 \alpha _1^2 \lambda _1-12 \alpha _2^2 \lambda _1-\frac{13}{4}
   \alpha _3^2 \lambda _1-15 \alpha _1 \alpha _3 \lambda _1 \nonumber \\
   && +72 h_t^4 \lambda _2-\frac{10}{3} g_{BL}^2 h_t^2 \lambda _2-160 g_S^2 h_t^2
   \lambda _2+60 \alpha _1^2 \lambda _2-240 \alpha _2^2 \lambda _2 \nonumber \\
   && -3 \alpha _3^2 \lambda _2+976 \lambda _1^2 \lambda _2+60 \alpha _1 \alpha _3
   \lambda _2+288 h_t^2 \lambda _1 \lambda _2-60 h_t^4 \lambda _3 \nonumber \\
   && +\frac{5}{3} g_{BL}^2 h_t^2 \lambda _3+80 g_S^2 h_t^2 \lambda _3-30
   \alpha _1^2 \lambda _3+24 \alpha _2^2 \lambda _3+\frac{3}{2} \alpha _3^2 \lambda _3 -664 \lambda _1^2 \lambda _3 \nonumber \\
   && -4480 \lambda _2^2 \lambda
   _3-30 \alpha _1 \alpha _3 \lambda _3-192 h_t^2 \lambda _1 \lambda _3
    +576 h_t^2 \lambda _2 \lambda _3
   +2624 \lambda _1 \lambda _2 \lambda_3 \nonumber \\
   && -\frac{1}{16} g_R^4 \left(18 h_t^2-240 \alpha _1-120 \alpha _3-157 \lambda _1-572 \lambda _2+166 \lambda _3\right) \nonumber \\
   &&+g_R^2 \left(\frac{45}{4} \lambda _1 h_t^2-45 \lambda _2 h_t^2+\frac{45}{2} \lambda _3 h_t^2+24
   \alpha _1^2
    +\frac{7 \alpha _3^2}{2}+78 \lambda _1^2+864 \lambda _2^2 \right. \nonumber \\
   &&  +96 \lambda _3^2+108 \lambda _4^2 +24 \alpha _1 \alpha _3-144 \lambda _1
   \lambda _2+120 \lambda _1 \lambda _3-576 \lambda _2 \lambda _3 \bigg) \bigg] \,, \\
\label{eqn:LR:beta4}
\beta (\lambda_4) & \ = \ &
\frac{1}{16\pi^2} \bigg[
3\alpha_2 (2\alpha_1 + \alpha_3) + \frac32 h_t^4
-9 \lambda_4 (g_L^2 + g_R^2)
+ 48 (\lambda_1 + 2 \lambda_2 + \lambda_3) \lambda_4 + 12 \lambda_4 h_t^2 \bigg] \nonumber \\
&& + \frac{1}{(16\pi^2)^2} \bigg[
-72 \alpha _1 \alpha _2 \lambda _1-36 \alpha _2 \alpha _3 \lambda _1-144 \alpha _1 \alpha _2 \lambda _2-72 \alpha _2 \alpha _3 \lambda _2-72
   \alpha _1 \alpha _2 \lambda _3 \nonumber \\
   && -36 \alpha _2 \alpha _3 \lambda _3-30 \alpha _1^2 \lambda _4-120 \alpha _2^2 \lambda _4-\frac{9}{2} \alpha
   _3^2 \lambda _4-30 \alpha _1 \alpha _3 \lambda _4-48 \alpha _2^3 \nonumber \\
   && -15 \alpha _2 \alpha _3^2-36 \alpha _1^2 \alpha _2-36 \alpha _1 \alpha _2
   \alpha _3+g_{BL}^2 \left(48 \alpha _1 \alpha _2+24 \alpha _2 \alpha _3+\frac{5}{3} \lambda _4 h_t^2\right) \nonumber \\
   &&+\frac{3}{4} g_R^2
   \left(128 \alpha _1 \alpha _2+64 \alpha _3 \alpha _2+3 \lambda _4 \left(13 g_L^2+2 \left(5 h_t^2+48 \left(\lambda _1+2 \lambda _2+\lambda
   _3\right)\right)\right)\right) \nonumber \\
   && +\frac{45}{2} \lambda _4 g_L^2 h_t^2+80 \lambda _4 g_S^2 h_t^2-\frac{51}{8} \lambda _4 g_L^4+216 \lambda _1
   \lambda _4 g_L^2+432 \lambda _2 \lambda _4 g_L^2+216 \lambda _3 \lambda _4 g_L^2 \nonumber \\
   && +\left(30 \alpha _2+\frac{37 \lambda _4}{8}\right) g_R^4-36
   \lambda _4 h_t^4-288 \lambda _1 \lambda _4 h_t^2-576 \lambda _2 \lambda _4 h_t^2-288 \lambda _3 \lambda _4 h_t^2 \nonumber \\
   && -1248 \lambda _4^3-1064
   \lambda _1^2 \lambda _4-4992 \lambda _2^2 \lambda _4-1088 \lambda _3^2 \lambda _4 \nonumber \\
   && -3456 \lambda _1 \lambda _2 \lambda _4-1888 \lambda _1
   \lambda _3 \lambda _4-3840 \lambda _2 \lambda _3 \lambda _4 \bigg] \,, \\
\label{eqn:LR:beta5}
\beta (\rho_1) & \ = \ &
\frac{1}{16\pi^2} \bigg[
3 \left( 3 g_R^4 + 4 g_R^2 g_{BL}^2 + 2 g_{BL}^4 \right)
+ 2 \left( 2\alpha_1^2 + 8 \alpha_2^2 + 2\alpha_1 \alpha_3 + \alpha_3^2 \right) \nonumber \\
&& + 28 \rho_1^2 + 16 \rho_2^2
 -12 \rho_1 \left( 2g_R^2 + g_{BL}^2 \right) + 16 \rho_1 \rho_2 \bigg] \nonumber \\
&&  + \frac{1}{(16\pi^2)^2} \bigg[
-\frac{2}{3} \left(2 g_{BL}^4 \left(233 g_R^2-170 \rho _1-60 \rho _2\right)  +4 g_{BL}^2 \left(3 \left(-17 \rho _1+20
   \rho _2\right) g_R^2 \right. \right. \nonumber \\
   &&
   + \left. \left. 118 g_R^4
   -6 \left(11 \rho _1^2+4 \left(2 \rho _2\right) \rho _1
   +8 \rho _2^2\right)\right)
    +196   g_{BL}^6 \nonumber \right. \\
   &&
   +3 \left(20 \alpha _1^2 \rho _1+20 \alpha _3 \alpha _1 \rho _1+80 \alpha _2^2 \rho _1
   +11 \alpha _3^2 \rho _1
   +8 \alpha_1^3+12 \alpha _3 \alpha _1^2 \right. \nonumber \\
   &&
   +96 \alpha _2^2 \alpha _1+14 \alpha _3^2 \alpha _1+5 \alpha _3^3+48 \alpha _2^2 \alpha _3
    -6 \left(2 \alpha
   _1^2+2 \alpha _3 \alpha _1+8 \alpha _2^2+\alpha _3^2\right) g_L^2 \nonumber \\
   && \left.  +6 \left(2 \alpha _1^2+2 \alpha _3 \alpha _1+8 \alpha _2^2+\alpha
   _3^2\right) h_t^2
    +192 \rho _1^3+160 \rho _2^3+312 \rho _1 \rho _2^2+176 \rho _1^2 \rho _2\right) \nonumber \\
   &&
   -2 \left(30 \alpha _1+15 \alpha _3+154 \rho_1+96 \rho _2\right) g_R^4
    -3 \left(12 \alpha _1^2+12 \alpha _3 \alpha _1+48 \alpha _2^2 \right. \nonumber \\
   && \left. \left. +3 \alpha _3^2+176 \rho _1^2+80 \rho _2^2+128 \rho
   _1 \rho _2\right) g_R^2+33 g_R^6\right) \bigg] \,, \\
\label{eqn:LR:beta6}
\beta (\rho_2) & \ = \ &
\frac{1}{16\pi^2} \bigg[
3 g_R^2 \left( g_R^2 - 4 g_{BL}^2 \right)
- \alpha_3^2 + 12 \rho_2 (2\rho_1 + \rho_2)
-12 \rho_2 \left( 2g_R^2 + g_{BL}^2 \right) \bigg] \nonumber \\
&& +\frac{1}{(16\pi^2)^2} \bigg[
\frac{2}{3} \left(2 g_{BL}^4 \left(143 g_R^2+50 \rho _2\right)+4 g_{BL}^2 \left(9 \left(-4 \rho _1+7 \rho _2\right)
   g_R^2  +73 g_R^4 \right. \right. \nonumber \\
   && \left. +18 \left(2 \rho _1-\rho _2\right) \rho _2\right)
   -3 \left(-4 \alpha _3^2 \rho _1-\alpha _3^2 \rho _2+20 \alpha _1
   \alpha _3 \rho _2+20 \alpha _1^2 \rho _2
     \right. \nonumber \\
   && \left. +80 \alpha _2^2 \rho _2
   -2 \alpha _3^3
   -4 \alpha _1 \alpha _3^2+3 \alpha _3^2 g_L^2-3 \alpha _3^2
   h_t^2+8 \rho _2^3+224 \rho _1 \rho _2^2+224 \rho _1^2 \rho _2\right) \nonumber \\
   && \left. +4 \left(9 \rho _1+8 \rho _2\right) g_R^4+144 \rho _2 \left(2 \rho
   _1+\rho _2\right) g_R^2-119 g_R^6\right) \bigg] \,, \\
\label{eqn:LR:beta7}
\beta (\alpha_1) & \ = \ &
\frac{1}{16\pi^2} \bigg[
6 g_R^4
+ 4 \alpha_1^2 + 16 \alpha_2^2 + \alpha_3^2
+ 4 \alpha_1 (5 \lambda_1 + 2\lambda_3)
+ 48 \alpha_2 \lambda_4
+ 8 \alpha_3 (\lambda_1 + \lambda_3) \nonumber \\
&& + 8 \alpha_1 (2\rho_1 + \rho_2)
+ 2 \alpha_3 (3\rho_1 + 4 \rho_2)
 - \frac32 \alpha_1 ( 3 g_L^2 + 11g_R^2 + 4 g_{BL}^2 )
+ 6 \alpha_1 h_t^2 \bigg] \nonumber \\
&& + \frac{1}{(16\pi^2)^2} \bigg[
   \frac{245 g_R^6}{6}
   -30   g_R^4 g_{BL}^2
   -2  g_R^2 g_{BL}^2 \left(6 h_t^2-10
   \alpha _1+19 \alpha _3\right)
    \nonumber \\
   && -\frac{3}{16} g_L^4 \left(\alpha _1-40 \alpha _3\right)
   +g_{BL}^4 \left(-4 h_t^2+\frac{268 \alpha _1}{3}+30 \alpha _3\right) \nonumber \\
   && +g_R^4 \left( -6 h_t^2 +\frac{3647}{48} \alpha _1 +50 \alpha _3 +100 \lambda _1 +40 \lambda _3
   +80 \rho _1 +40 \rho _2 \right)  \nonumber \\
    && -\frac{3}{8} g_L^2 \left(60 g_R^4-\left(15
   \alpha _1+4 \alpha _3\right) g_R^2-2 \left(15 \alpha _1 h_t^2+4 \alpha _1^2+16 \alpha _2^2+\alpha _3^2 \right. \right. \nonumber \\
   && \left. \left.   +64 \alpha _3 \lambda _1+64 \alpha _3
   \lambda _3+32 \alpha _1 \left(5 \lambda _1+2 \lambda _3\right)+384 \alpha _2 \lambda _4\right)\right)\nonumber \\
   && +g_R^2 \left(
   11 \alpha _1^2 +44 \alpha _2^2 + \frac{11}{4} \alpha _3^2 +\frac{45}{4} h_t^2 \alpha _1
    +120   \alpha _1 \lambda _1
    +60 \alpha _3 \lambda _1 \right. \nonumber \\
   && +48 \alpha _1 \lambda _3
   +24 \alpha _3 \lambda _3
    +288 \alpha _2 \lambda_4
   +256 \alpha _1 \rho _1
   +108 \alpha _3 \rho _1
   +128 \alpha _1 \rho _2
   +104 \alpha _3 \rho _2  \bigg) \nonumber \\
   && +g_{BL}^2 \bigg( 4 \alpha _1^2+16 \alpha _2^2+\alpha _3^2 +48 \alpha _3 \rho _1+64 \alpha _3 \rho _2 \nonumber \\
   && \left. +\alpha _1
   \left(\frac{5 h_t^2}{6}+32 \left(4 \rho _1+2 \rho _2\right)\right)\right)
   -64   \alpha _1 \alpha _3 \rho _2-80 \alpha _1 \rho _1 \rho _2-64 \alpha _3 \rho _1 \rho _2\nonumber \\
   &&    -15 \alpha _1^3-6
   \alpha _3^3
   -12 h_t^2 \alpha _1^2
   -48 h_t^2 \alpha _2^2
    -172 \alpha _1 \alpha _2^2
   -3 h_t^2 \alpha _3^2-\frac{45}{4} \alpha _1 \alpha _3^2
   \nonumber \\
   &&
   -100   \alpha _1 \lambda _1^2-32 \alpha _3 \lambda _1^2-960 \alpha _1 \lambda _2^2
    -768 \alpha _3 \lambda _2^2-160 \alpha _1 \lambda _3^2-128
   \alpha _3 \lambda _3^2\nonumber \\
   &&-240 \alpha _1 \lambda _4^2-144 \alpha _3 \lambda _4^2-80 \alpha _1 \rho _1^2
   -24 \alpha _3 \rho _1^2-120 \alpha _1
   \rho _2^2-96 \alpha _3 \rho _2^2 \nonumber \\
   && -18 h_t^4 \alpha _1+40 g_S^2 h_t^2 \alpha _1
   -7   \alpha _1^2 \alpha _3-80 \alpha _2^2 \alpha _3
   -120 \alpha _1^2
   \lambda _1-224 \alpha _2^2 \lambda _1 \nonumber \\
   &&  -30 \alpha _3^2 \lambda _1-120 h_t^2 \alpha _1 \lambda _1-48 h_t^2 \alpha _3 \lambda _1-64 \alpha _1
   \alpha _3 \lambda _1-768 \alpha _2^2 \lambda _2
    -48 \alpha _1^2 \lambda _3\nonumber \\
   &&
   -320 \alpha _2^2 \lambda _3-12 \alpha _3^2 \lambda _3-48 h_t^2
   \alpha _1 \lambda _3-48 h_t^2 \alpha _3 \lambda _3-64 \alpha _1 \alpha _3 \lambda _3
    -80 \alpha _1 \lambda _1 \lambda _3 \nonumber \\
   &&   -64 \alpha _3
   \lambda _1 \lambda _3-288 h_t^2 \alpha _2 \lambda _4
   -576 \alpha _1 \alpha _2 \lambda _4-192 \alpha _2 \alpha _3 \lambda _4-576 \alpha _2
   \lambda _1 \lambda _4 \nonumber \\
   && -1152 \alpha _2 \lambda _2 \lambda _4
   -576 \alpha _2 \lambda _3 \lambda _4
   -96 \alpha _1^2 \rho _1-384 \alpha
   _2^2 \rho _1-24 \alpha _3^2 \rho _1-48 \alpha _1 \alpha _3 \rho _1 \nonumber \\
   && -48 \alpha _1^2 \rho _2-192 \alpha _2^2 \rho _2
   -12 \alpha _3^2 \rho _2
\bigg] \,, \\
\label{eqn:LR:beta8}
\beta (\alpha_2) & \ = \ &
\frac{1}{16\pi^2} \bigg[
- \frac32 \alpha_2 ( 3 g_L^2 + 11g_R^2 + 4 g_{BL}^2 )
+ 6 (2 \alpha_1 + \alpha_3) \lambda_4
+ 4 \alpha_2 ( \lambda_1 + 12 \lambda_2 + 4 \lambda_3 ) \nonumber \\
&& + 8 \alpha_2 ( 2\rho_1 + \rho_2 )
+ 4 \alpha_2 ( 2\alpha_1 + \alpha_3 )
+ 6 \alpha_2 h_t^2 \bigg] \nonumber \\
&& + \frac{1}{(16\pi^2)^2} \bigg[
 -\frac{243}{16} \alpha _2 g_L^4
  +\frac{45}{8}  \alpha _2 g_L^2  g_R^2
    + g_R^4 \left( \frac{2927}{48} \alpha _2 +60 \lambda _4 \right)
 +\frac{268}{3} \alpha _2 g_{BL}^4
  \nonumber \\
 &&   +40 \alpha _2    g_S^2 h_t^2
   +\frac{3}{8} g_L^2
    \left(2 \left(4 \left(\alpha _2 \left(\alpha _3+8 \lambda _1+96 \lambda _2+32 \lambda _3\right) \right. \right. \right. \nonumber \\
    && \left. \left. \left. +12 \alpha _3 \lambda
   _4+2 \alpha _1 \left(\alpha _2+12 \lambda _4\right)\right)+15 \alpha _2 h_t^2\right)\right) \nonumber \\
      &&  + g_R^2 \bigg(
   24 \alpha _2 \lambda _1
   +288 \alpha _2 \lambda _2
   +96 \alpha _2 \lambda _3 g_R^2
   +72    \alpha _1 \lambda _4
   +36 \alpha _3 \lambda _4
   +256 \alpha _2 \rho _1 \nonumber \\
  && \left. +128 \alpha _2 \rho _2
      +22 \alpha _1 \alpha _2
      +11 \alpha _2 \alpha _3
      +\frac{45}{4} \alpha _2  h_t^2 \right) \nonumber \\
   &&+\frac{1}{6} \alpha _2   g_{BL}^2 \left(24 \left(2 \alpha _1+\alpha _3+32 \rho _1+16 \rho _2\right)+120 g_R^2+5 h_t^2\right) \nonumber \\
   && -36 \alpha _2 \lambda _1^2+192 \alpha _2 \lambda _2^2-240 \alpha _2 \lambda _4^2-112 \alpha _1 \alpha _2 \lambda _1-56 \alpha _2 \alpha _3
   \lambda _1-384 \alpha _1 \alpha _2 \lambda _2 \nonumber \\
   && -192 \alpha _2 \alpha _3 \lambda _2-384 \alpha _2 \lambda _1 \lambda _2-160 \alpha _1 \alpha
   _2 \lambda _3-80 \alpha _2 \alpha _3 \lambda _3-144 \alpha _2 \lambda _1 \lambda _3 \nonumber \\
   && -384 \alpha _2 \lambda _2 \lambda _3-72 \alpha _1^2
   \lambda _4-288 \alpha _2^2 \lambda _4-30 \alpha _3^2 \lambda _4-72 \alpha _1 \alpha _3 \lambda _4-144 \alpha _1 \lambda _1 \lambda _4 \nonumber \\
   && -72
   \alpha _3 \lambda _1 \lambda _4-288 \alpha _1 \lambda _2 \lambda _4-144 \alpha _3 \lambda _2 \lambda _4-144 \alpha _1 \lambda _3 \lambda
   _4 \nonumber \\
   && -72 \alpha _3 \lambda _3 \lambda _4-80 \alpha _2 \rho _1^2-120 \alpha _2 \rho _2^2-192 \alpha _1 \alpha _2 \rho _1-96 \alpha _2 \alpha _3
   \rho _1-96 \alpha _1 \alpha _2 \rho _2 \nonumber \\
   && -48 \alpha _2 \alpha _3 \rho _2-80 \alpha _2 \rho _1 \rho _2-60 \alpha _2^3-\frac{49}{4} \alpha _2
   \alpha _3^2-43 \alpha _1^2 \alpha _2-43 \alpha _1 \alpha _2 \alpha _3
    -30 \alpha _2 h_t^4\nonumber \\
   && -12 h_t^2  \left( 2 \alpha _2 \lambda _1
      +24 \alpha _2 \lambda _2
   +8 \alpha _2 \lambda _3
   +6 \alpha _1 \lambda _4
   +3 \alpha _3 \lambda _4
   +2 \alpha _1 \alpha _2
   + \alpha _2 \alpha _3
   \right)
\bigg] \,, \\
\label{eqn:LR:beta9}
\beta (\alpha_3) & \ = \ &
\frac{1}{16\pi^2} \bigg[
4 \alpha_3^2
- \frac32 \alpha_3 \left( 3 g_L^2 + 11 g_R^2 + 4 g_{BL}^2 \right)
+ 8 \alpha_1 \alpha_3
+ 4 \alpha_3 (\lambda_1 -2 \lambda_3)\nonumber \\
&&
+ 4 \alpha_3 (\rho_1 - 2 \rho_2)
+ 6 \alpha_3 h_t^2 \bigg] \nonumber \\
&& + \frac{1}{(16\pi^2)^2} \bigg[
\frac{88}{3} \alpha _3 g_{BL}^4+\frac{1}{6} g_{BL}^2 \left(144 g_R^2 \left(4 \alpha _3+h_t^2\right) \right. \nonumber \\
   && \left. +\alpha _3 \left(24 \left(2
   \alpha _1+\alpha _3+8 \left(\rho _1-2 \rho _2\right)\right)+5 h_t^2\right)\right) \nonumber \\
   && -\frac{1}{48} \alpha _3 \left(-18 g_L^2 \left(16
   \alpha _1+8 \alpha _3+7 g_R^2+30 h_t^2+64 \lambda _1-128 \lambda _3\right) \right. \nonumber \\
   && \left.  -12 g_R^2 \left(88 \alpha _1+44 \alpha _3+45 h_t^2+160 \rho
   _1-320 \rho _2\right) \right. \nonumber \\
   &&  -12 \left(-448 \alpha _1 \lambda _1-224 \alpha _3 \lambda _1+128 \alpha _1 \lambda _3+64 \alpha _3 \lambda _3-768
   \alpha _2 \lambda _4 \right. \nonumber \\
   &&  -384 \alpha _1 \rho _1-192 \alpha _3 \rho _1+128 \alpha _1 \rho _2+64 \alpha _3 \rho _2-124 \alpha _1^2 \nonumber \\
   && -48 \alpha
   _2^2-29 \alpha _3^2-124 \alpha _1 \alpha _3+160 g_S^2 h_t^2
   +384 \lambda _3^2+192 \lambda _4^2 \nonumber \\
   &&   -48 \left(2 \alpha _1+\alpha _3+2 \lambda _1-4 \lambda _3\right) h_t^2-72
   h_t^4-144 \lambda _1^2+2304 \lambda _2^2 \nonumber \\
   &&  \left. \left. +192 \lambda _1 \lambda _3-128 \rho _1^2+288 \rho _2^2+192 \rho
   _1 \rho _2\right)
    +729 g_L^4+1153 g_R^4\right)
\bigg] \,, \\
\label{eqn:LR:beta10}
\beta (h_t) & \ = \ &
\frac{1}{16\pi^2} \bigg[
- h_t \left( 8 g_S^2 + \frac94 g_L^2 + \frac94 g_R^2 + \frac16 g_{BL}^2 \right) + 5 h_t^3 \bigg] \nonumber \\
&& + \frac{1}{(16\pi^2)^2} \bigg[
-\frac{1}{144} h_t \bigg(-g_{BL}^2 \left(4 \left(-8 g_S^2+31 h_t^2\right)+27 g_L^2+27 g_R^2\right) \nonumber \\
   && -97 g_{BL}^4+3
   \left(9 g_L^2 \left(27 g_R^2-48 g_S^2-86 h_t^2\right) \right. \nonumber \\
   &&  +2 \left(-9 g_R^2 \left(24 g_S^2+43 h_t^2\right)-992 g_S^2 h_t^2+78 g_R^4+2592 g_S^4 \right. \nonumber \\
   && +3
   \left(-12 \alpha _1^2-48 \alpha _2^2-9 \alpha _3^2-12 \alpha _1 \alpha _3+64 \left(2 \lambda _1-\lambda _3\right) h_t^2 \right. \nonumber \\
   &&  \left. \left. \left. +136 h_t^4-80
   \lambda _1^2-768 \lambda _2^2-128 \lambda _3^2-192 \lambda _4^2-64 \lambda _1 \lambda _3\right)\right)+252 g_L^4\right) \bigg)
\bigg] \,.
\end{eqnarray}


To see how the fermions get their masses in the LRSM, we write down the Yukawa Lagrangian:
\begin{eqnarray}
\label{eqn:Lyukawa}
{\cal L}_Y \ & = & \
h^q \bar{Q}_{L} \Phi Q_{R} +
\tilde{h}^q_{} \bar{Q}_{L} \tilde{\Phi} Q_{R} +
h^\ell_{} \bar{\psi}_{L} \Phi \psi_{R} +
\tilde{h}^\ell_{} \bar{\psi}_{L} \tilde{\Phi} \psi_{R} \nonumber \\
&& + f_{R} \psi_{R}^{\sf T} C i\tau_2 \Delta_R \psi_{R}  ~+~ {\rm H.c.}
\end{eqnarray}
where $\tilde{\Phi}=\sigma_2\Phi^*\sigma_2$ ($\sigma_2$ being  the  second Pauli matrix) and $C=i\gamma_2\gamma_0$ is the charge conjugation operator ($\gamma_{\mu}$ being the Dirac matrices). After symmetry breaking, the quark and charged lepton masses are given by the generic formulas $M_u = h^q \kappa + \tilde{h}^q \kappa'$ for up-type quarks, $M_{d} = h^q \kappa' + \tilde{h}^q \kappa$ for down-type quarks, and similarly for the charged leptons, where we have neglected CP violation in the fermion matrices. To account for the SM fermion hierarchy, we set $\kappa'/\kappa \simeq m_b/m_t \simeq 1/60$, then the top and bottom quark masses are respectively
\begin{eqnarray}
m_t \ \simeq \ h^q_{33} \kappa \simeq h_{33}^q v_{\rm EW} \,, \quad
m_b \ \simeq \ h^q_{33} \kappa^\prime + \tilde{h}^q_{33} \kappa \,,
\end{eqnarray}
with $h^q_{33}$ and $\tilde{h}^q_{33}$ the $(3,\,3)$ elements of the $h^q$ and $\tilde{h}^q$ matrices. It is expected that for the bottom quark mass $\tilde{h}^q_{33} \ll h^q_{33} \sim {\cal O} (1)$. With the first two generation quarks much lighter than the third generation in the SM, we consider only the RG running of $h_t = h^q_{33}$ in  the quark sector, as shown in Eq.~(\ref{eqn:LR:beta10}).

In the lepton sector, the tauon mass $m_\tau \simeq h^\ell_{33} \kappa^\prime + \tilde{h}^\ell_{33} \kappa$ ($h^\ell_{33}$ and $\tilde{h}^\ell_{33}$ are respectively the $(3,\,3)$ elements of the $h^\ell$ and $\tilde{h}^\ell$ matrices), which is closely related to the Dirac mass matrix  for neutrinos $m_D = h^{\ell} \kappa + \tilde{h}^{\ell} \kappa'$. The elements $h^\ell_{33}$ and $\tilde{h}^\ell_{33}$ cannot be very large for TeV-scale RHNs, or we need fine-tuning or large cancellation in fitting the charged lepton masses and the tiny neutrino masses. Thus we have neglected also the matrices $h^\ell$ and $\tilde{h}^\ell$ in the $\beta$-functions above. For the RH scale $v_R \gtrsim 10$ TeV, as implied by the scalar perturbativity constraints in Figs.~\ref{fig:LRSM3} and \ref{fig:LRSM4}, if the RHNs are all the TeV-scale, say $M_N \simeq 1$ TeV, the Yukawa coupling $f_R \sim M_N / v_R \lesssim 0.1$, and we do not include it either in the $\beta$-functions above.

\end{document}